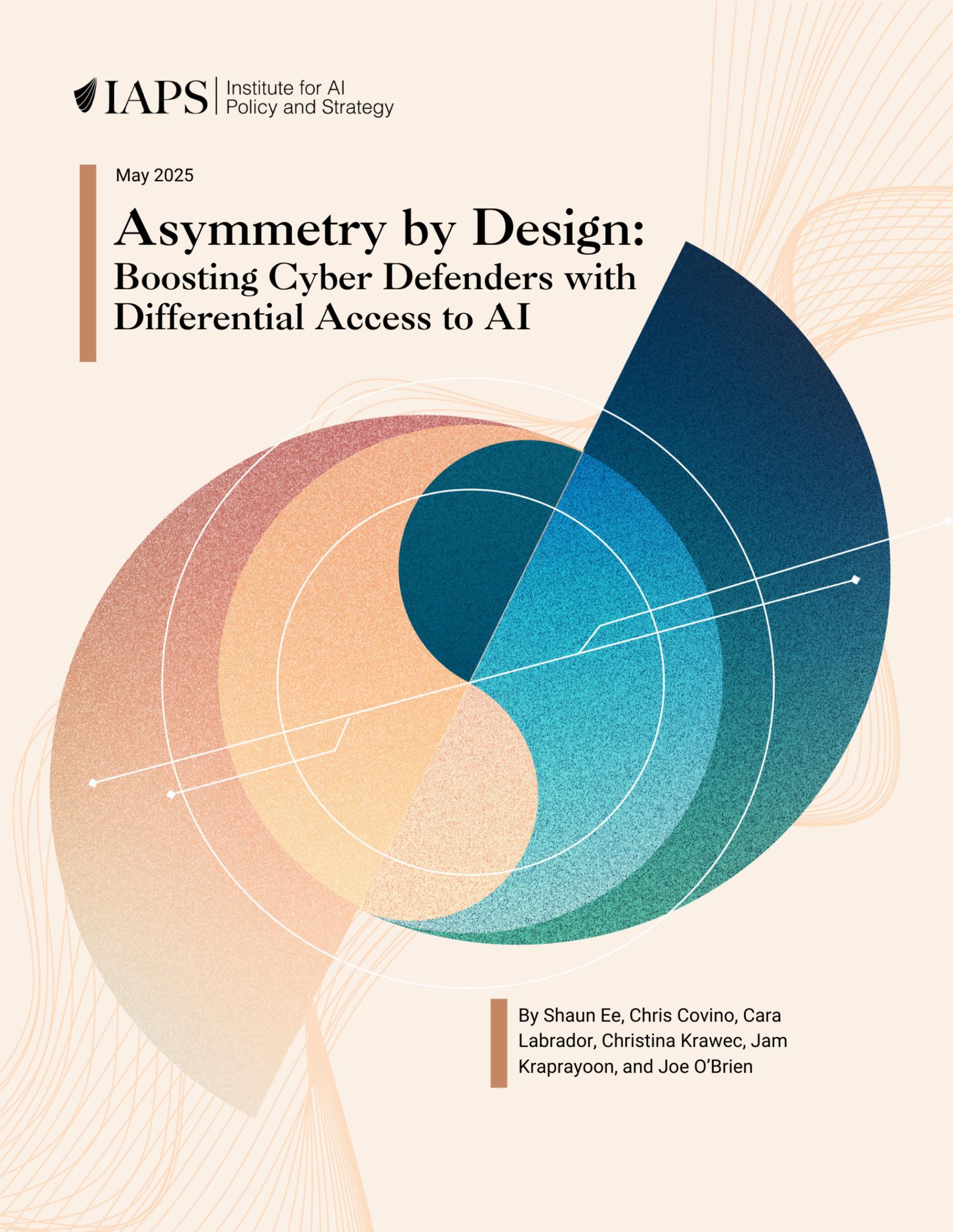

IAPS | Institute for AI Policy and Strategy

May 2025

# Asymmetry by Design:
## Boosting Cyber Defenders with Differential Access to AI

By Shaun Ee, Chris Covino, Cara Labrador, Christina Krawec, Jam Kraprayoon, and Joe O'Brien

# Table of Contents

# Executive Summary

The rise of Artificial Intelligence (AI) models will significantly impact the cybersecurity ecosystem. From vulnerability discovery to network penetration, AI-enabled cyber (AIxCyber) capabilities are already enhancing cybersecurity applications and operations, providing tremendous benefits for defenders.

Unfortunately, widespread access also brings significant risk. Malicious actors will leverage AI to infiltrate networks, deploy ransomware, and conduct other cyber attacks. Many defenders lack the resources and expertise to leverage AI cybersecurity applications or integrate AI into their security operations. These organisations already face serious challenges, as recent major American telecommunications breaches and daily ransomware attacks on schools and hospitals demonstrate. Without a strategic rollout of AI cybersecurity capabilities that favours defenders, we risk a cyber landscape that shifts disproportionately toward attackers, worsening existing security gaps and potentially rendering today's cyber defences obsolete.

**"Differential access" is a strategy to tilt the cybersecurity balance toward defense by shaping access to advanced AI-powered cyber capabilities. The goal is to provide cyber defenders an asymmetric advantage over malicious attackers.** Here we introduce three possible differential access approaches:

- **Promote Access:** Prioritize widespread adoption of AIxCyber capabilities through open access and active promotion. This approach suits lower-risk capabilities, focusing on innovation and diffusion among strategically important defenders.
- **Manage Access:** Balance opportunities and risks of medium-risk capabilities through controlled distribution and prioritization of certain defenders. This approach combines access restriction with targeted promotion.
- **Deny by Default:** Restrict higher-risk cybersecurity capabilities to select defenders. This approach prioritizes risk mitigation while still providing defensive benefits through strategic diffusion.

These approaches form a continuum, becoming progressively more restrictive as AIxCyber capabilities increase in power. **However, a key principle across all approaches is defender access—even in the most restrictive scenarios, developers should advantage cyber defenders.**

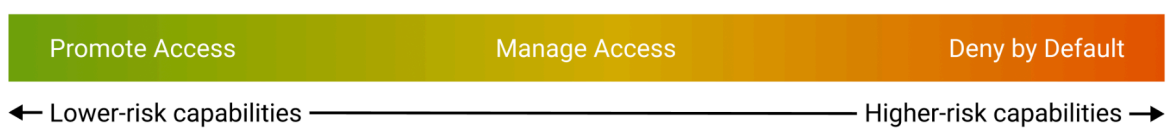

**This report provides a process to help developers choose and implement one of the three differential access approaches.** This process includes considerations based on a model's cyber capabilities, a defender's maturity and role, and strategic and technical implementation details.

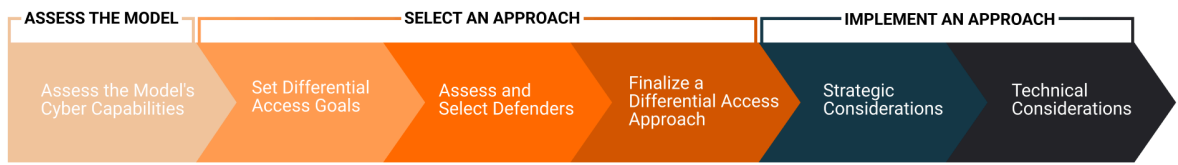

1. **Assess the model's cyber capabilities:** Evaluate the model's current or potential AIxCyber capabilities. Assess these capabilities for their defensive potential and risk of misuse. This will further require forecasting future capabilities, understanding use cases, and assessing how downstream developers may build upon the model's capabilities (i.e., fine-tuning, scaffolding, etc.)
2. **Set differential access goals:** Define specific goals and objectives for the differential access approach. Establishing clear goals helps identify which defenders will be the highest priority in pursuing these goals.
3. **Assess defender levels and select defenders:** Evaluate potential defenders based on their defender level—defined by both their ability to securely leverage AIxCyber capabilities and their criticality to the cybersecurity ecosystem or broader society.
4. **Finalize a differential access approach**: Determine a differential access approach—Promote Access, Manage Access, or Deny by Default—based on the model capabilities levels, defender assessments, and differential access goals.
5. **Strategic considerations:** Consider organizational strategies, policies, and other actions to implement the selected differential access approach.
6. **Technical considerations:** Consider the appropriate technical controls required to implement the differential access approach.

Finally, we present four example schemes that developers can reference. These schemes, though not exhaustive, demonstrate how differential access provides value across various capability and defender levels. **We believe differential access approaches work best when tied to specific purposes and goals,** such as: (1) addressing concrete threat scenarios like attacks on Critical National Infrastructure (CNI), or (2) mitigating specific concerns about advancing AI capabilities. For example, by countering AI-enabled vulnerability discovery and exploitation with AI systems that support secure software development and rapid patching.

| Name of Scheme | Type of Access | Description *(with number of users mentioned as a heuristic for degree of access allowed)* |
|---|---|---|
| Scheme A: Accelerator for CNI Cybersecurity Innovators | Promote Access | Early access and policy incentives (e.g., financial) for large number of startups and established product vendors to incentivize AI-based solutions for CNI security (e.g., better threat detection, identifying misconfigurations) |
| Scheme B: Dual-Use Authorization for Security Researchers | Manage Access | Expanded access to dual-use capabilities (e.g., exploit generation) for large number of white-hat hackers to improve the security of the open-source community, using technical infrastructure to manage and monitor their access |
| Scheme C: Rapid Response Force of Keystone Defenders | Manage Access | Foundation model developers convene a smaller, trusted group of keystone defenders to reduce vulnerabilities in software and improve threat detection, focusing on rapid experimentation and prototyping of new AI capabilities |
| Scheme D: High-Capability Adversarial Testing as a Service | Deny by Default | Foundation model owner/operator (e.g., government) tightly controls access to highly capable system (access for very few users) but conducts high-end penetration testing and red teaming for other actors, providing automated "nation-state attack emulation as a service" |

**Call to action:** We encourage readers—whether foundation model developers, policymakers, or researchers—to apply our differential access framework to their specific challenges. The exemplar schemes presented are not comprehensive, and many issues—from supply chain security for frontier AI developers to cybersecurity for the defense sector—remain unaddressed. We welcome others to build upon and improve this work.

# 1 | Differential Access Approaches

As AI systems become increasingly proficient at performing advanced cyber operations, policymakers and companies must guard against harmful applications of these capabilities—not just by restricting access to prevent misuse, but by actively empowering defenders. Current approaches are insufficient, as developers often determine access based on the capability's negative use cases without systematically evaluating whether defenders can safely adopt that capability. As a result, some organizations are needlessly deprived of AI tools that would help tilt the offense-defense balance in favor of cyber defenders.

We propose three differential access approaches with differing levels of access and oversight: Promote Access, Manage Access, and Deny by Default. Each approach is tailored to different scenarios based on an evaluation of an AI model's capability level and the defender's maturity and criticality. Typically, the more capable the model, the more mature a defender must be in order to gain access to that model, though developers may be able to expand access to additional defenders by implementing organizational strategies and technical controls.

These approaches are intended to be flexible and adaptable, providing developers with a decision-making framework and key considerations for access management that allow them to strike the right balance between empowering defenders and thwarting attackers.

This paper focuses on the cyber capabilities of **foundation models** and their **derivative cyber products or systems**. Foundation models are broad-based AI systems trained on diverse datasets that serve as a versatile platform for adaptation (e.g., GPT-4, LLaMA, and DeepSeek R1), while derivative products refer to specialized applications built on top of foundation models to address specific use cases (e.g., Google's Big Sleep vulnerability discovery agent[1]).

> ### SCENARIO: What might differential access look like?
>
> Imagine an AI developer creates a model able to identify complex vulnerabilities in code. Rather than restricting it entirely or making it widely available, the developer decides to strategically release to select defenders. They first set specific goals to reduce

---

[1] Google Project Zero, "Project Zero."

ASYMMETRY BY DESIGN | 6

> vulnerabilities in widely-used open-source libraries. Next, they assess possible cyber defenders to help them reach this goal, selecting security researchers and open-source developers. Once the defenders are selected, they finalize their approach and plan.
>
> Working with downstream developers, they provide vetted security researchers and open-source developers with early access to an AI agent that can identify vulnerabilities in open-source software before broader release. This strategic approach strengthens open-source security while preventing potential misuse of advanced AI capabilities.

## 1.1 | The Case for Differential Access

We define "differential access" as **strategically shaping access to AI systems with advanced cyber capabilities to advantage cyber defenders**. Implemented successfully, differential access approaches can restrict malicious actors from accessing AI-enabled cyber (AIxCyber) capabilities while also ensuring legitimate users can leverage them for defensive purposes.

While AI developers already implement some forms of differential access—such as safeguards targeted at preventing malware development—**existing measures primarily focus on preventing misuse rather than promoting adoption for defenders**. When assessing whether a cybersecurity capability should be made widely available, developers often make a determination based on the capability's negative use cases without evaluating whether critical defenders can safely adopt that capability.[2] Further, even if AIxCyber capabilities are available, many defenders may lack the resources and expertise to deploy them.

**As a result, cyber defenders are often deprived of AIxCyber capabilities** that would allow them to better safeguard users, systems, supply chains, and infrastructure. Access restrictions cannot fully prevent adversaries from gaining access to similar capabilities via tapping open-source models just behind the frontier, stealing advanced models, or even developing their own capabilities (as Chinese intelligence services might). **Denying all cyber defenders access to dual-use AIxCyber capabilities can create security risks**, creating an "offensive overhang" where organizations become increasingly vulnerable to adversaries that are offensively deploying similar capabilities and handicapping cybersecurity researchers who could otherwise bolster organizations' cybersecurity.

---

[2] For example, Anthropic's Responsible Scaling Policy evaluates the potential harm and misuse risk of specific cybersecurity capabilities to determine appropriate access limitations for their models. Source: "Responsible Scaling Policy."

As developers grapple with increasingly cyber-capable models and systems, **they need a more strategic and structured approach to differential access—one that prioritizes defensive access and is informed by the developer's goals, the model's capabilities, and the unique role different cyber defenders play within the cybersecurity ecosystem.**

## 1.2 | Overview and Selection of Differential Access Approaches

### Three Approaches to Differential Access

This report presents three differential access approaches that developers can use to manage AI systems with advanced cybersecurity capabilities in ways that balance innovation and security (see Sections 1.3-1.5 for further detail). These approaches exist along a spectrum from open to restricted access, and informed by a model's capabilities.

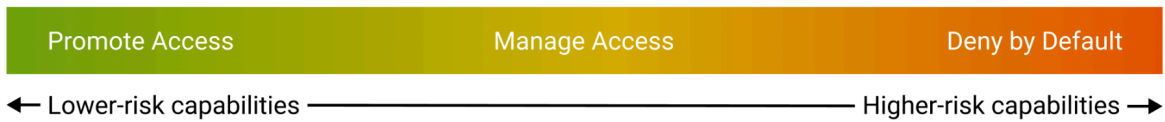

*The above diagram is illustrative, and the approach chosen may also vary by specific defender.*

- **Promote Access:** Prioritize expanding defender access to cybersecurity capabilities through open access (including open sourcing models) or actively promoting adoption to all or specific defenders. This approach is ideal for systems with low-risk capabilities or when capabilities can support a specific defender or address a cybersecurity gap. Providing open access encourages not only adoption but also innovation and productization. In addition to open access, developers can more actively increase adoption through early access programs, financial incentives, or collaborations with governments and third-party service providers. Higher-level or offensive-dominant capabilities may heighten the risk of misuse, as malicious actors may exploit openly available resources. In these situations, developers can consider selective access promotion through targeted incentives, trusted third-party services, and other mechanisms.
- **Manage Access:** Managed access requires more precise control over who can access specific capabilities, allowing advanced capabilities to reach and benefit defenders while preventing potential misuse. This approach offers a middle ground between fully open and closed approaches. This is particularly suitable for medium- and higher-risk capabilities where balancing restriction and accessibility becomes critical. Naturally, this requires more careful consideration when selecting defenders, especially for higher-capability tools that could pose greater risks if misused. Controls here could include more robust know-your-customer requirements, coupled with fine-grained access controls. While this method may limit malicious use, it has the drawback of being more administratively and technically complex, possibly requiring novel technical measures to implement effectively

ASYMMETRY BY DESIGN | 8

(see Section 5.2). It may also limit widespread adoption, so developers may want to prioritize defenders whose security work benefits others, such as software suppliers or third-party cybersecurity providers.
- **Deny by Default:** A Deny by Default approach only provides access to a select few defenders. This approach is most appropriate for the highest-risk capabilities, e.g., capabilities that broadly offer nation-state-level offensive advantages or where misuse risks are significant. Deny by Default requires comprehensive security protocols, rigorous vetting processes, and continuous monitoring. This approach demands thorough evaluation of potential users against stringent criteria. While maximizing security through strict limitations, this method significantly constrains broader defensive benefits that might come from wider adoption. Developers might mitigate this limitation by establishing secure environments where vetted researchers or defenders can utilize capabilities under supervision, or by creating derived tools with reduced capabilities for broader distribution. In its most extreme form, this approach could entail barring all but the model's developers and governments' access to these capabilities.

## Selecting the Appropriate Approach

The selection of approach depends on systematically evaluating three key factors: model capabilities, defender characteristics, and strategic objectives. This report provides a process for selecting and implementing differential access approaches. We encourage developers to follow this process as they consider differential access programs for a foundational model, specific model's capabilities, or even derivative products.

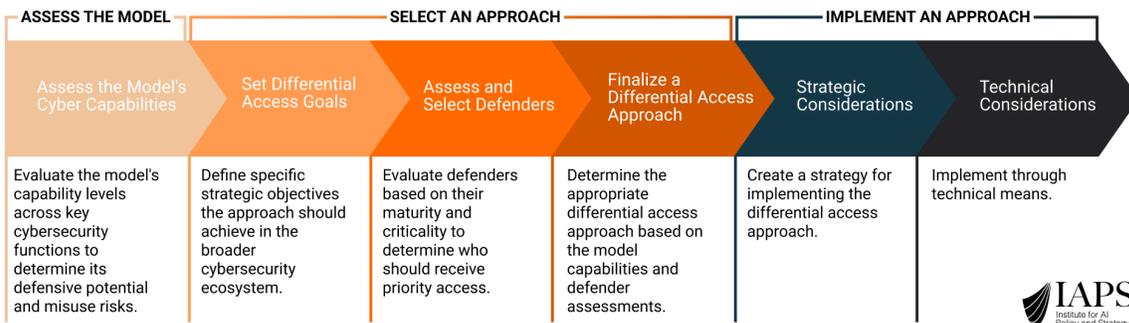

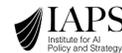

1. **Assess the model's cyber capabilities:** Evaluate the model's current or potential AIxCyber capabilities. Assess these capabilities for their defensive potential and risk of misuse. This will further require forecasting future capabilities, understanding use cases, and assessing how downstream developers may build upon the model's capabilities (e.g., fine-tuning, scaffolding).
2. **Set differential access goals:** Define specific goals and objectives for the differential access approach. Establishing clear goals helps identify which defenders will be the highest priority in pursuing these goals.

3. **Assess defender levels and select defenders:** Evaluate potential defenders based on their defender level, which is their ability to securely leverage AIxCyber capabilities and their criticality to the cybersecurity ecosystem or broader society.
4. **Finalize a differential access approach:** Determine the appropriate differential access approach—Promote Access, Manage Access, or Deny by Default—based on the model capability levels, defender assessments, and differential access goals.
5. **Strategic considerations:** Consider organizational strategies, policies, and other actions to implement the selected differential access approach.
6. **Technical considerations:** Consider the appropriate technical controls required to implement the differential access approach.

Ultimately, differential access is about balancing access and restrictions by considering the capability and the defenders. For example, the below figure illustrates how a developer may want to pursue a middle-of-the-road Manage Access approach for a potentially high-risk system. This Manage Access approach may create opportunities for medium- and high-level defenders to gain controlled access, e.g., cybersecurity service providers, critical infrastructure, or software suppliers. At the same time, it would restrict access to low-criticality and low-maturity defenders, such as small businesses or the general public.

## Example: Selecting differential access approach based on capability level and defender maturity

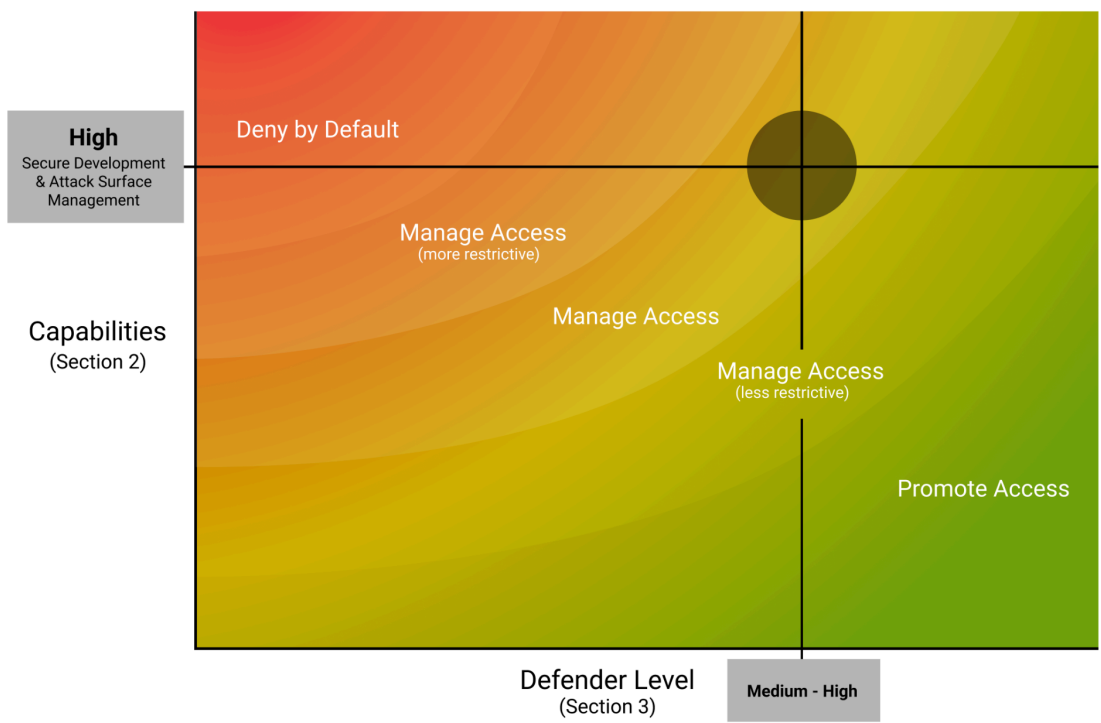

It is important to note that **these differential access approaches are intended to be flexible and adaptable**, providing developers with a decision-making framework and key considerations rather than rigid rules. The three approaches outlined—Promote Access, Manage Access, and Deny by Default—are not mutually exclusive. Developers may choose to blend elements from different approaches based on the specific context. The goal is to strike the right balance between empowering defenders and mitigating potential misuse or unauthorized access.

The remainder of Section 1 provides in-depth descriptions of the Promote Access, Deny by Default, and Manage Access approaches, including their primary benefits, drawbacks, and applicable scenarios. Sections 2-3 address the approach selection criteria in detail, helping developers identify Differential Access goals, model capability levels, and defender levels. Section 4 addresses organizational strategies for effective implementation, while Section 5 discusses technical infrastructure considerations. Finally, Section 6 provides a series of sample differential access schemes for key use cases that developers can consult when designing their own approach.

## 1.3 | Promote Access

The Promote Access approach prioritizes expanding defender access to a model or system's cybersecurity capabilities through passive or active efforts. At lower capability levels, this may include open-sourcing a model or even encouraging adoption among specific defenders through targeted incentives or other efforts. As a model or system's capabilities increase, this approach may become more selective, with promotion efforts specifically targeted to more mature and critical defenders. More active efforts to increase adoption will require assessing defenders' roles and barriers to adoption.

### Applicability

This approach is most appropriate in scenarios where AI systems exhibit relatively modest capability levels, such as those typically found in widely available open-source models. It is likely relevant to much of the current AI ecosystem, where foundation models do not yet pose widespread threats to public safety or national security, despite rapid advances in both general and cybersecurity-specific capabilities.

Promote Access strategies can remain viable even as AI capabilities advance. Open-source models will likely continue to improve in capabilities, even if the most advanced systems (frontier AI systems) remain proprietary. When open-source models already match or exceed a foundation model's capabilities, developers may prefer a Promote Access strategy to maximize adoption among cybersecurity defenders. However, this approach should, whenever possible, be paired with appropriate governance frameworks and misuse monitoring.

### Benefits and Drawbacks

- **Encourages innovation:** Diverse stakeholders—including individual researchers, think tanks, startups, and universities—can leverage this approach for experimentation and development of novel AIxCyber applications. Not only does this model foster independent innovation, but it supports cross-collaboration across nations and industries.
- **Promotes capabilities diffusion:** Democratizing access to AIxCyber capabilities has the potential to strengthen allies and partners outside of one's jurisdiction, though this can be a double-edged sword: without restrictions, malicious actors could easily exploit AI tools for cyberattacks, fraud, and other illicit activities.
- **Lack of oversight:** Insufficient oversight could result in unintended misuse, particularly by actors without proper training or skills, posing additional challenges for security and compliance.

## 1.4 | Manage Access

A Manage Access approach is more selective about granting access to advanced AIxCyber capabilities, balancing potential misuse risks with defensive benefits. Differential access schemes in this vein might take the form of tiered schemes that limit public access and restrict less mature users while still enabling defenders and the overall cyber ecosystem to benefit. The below table illustrates a high-level conceptual example of what such tiered access might look like.

### Sample Tiered Access Levels by Defender Characteristics

| Access Level | Who? (example) | Why? (example) | Type of Access Granted (example) |
|---|---|---|---|
| Tier 1 | Keystone Defenders (e.g., major OS and mobile developers, cybersecurity product vendors, software suppliers) | High maturity enables effective use of new technology; high security reduces risk of attackers gaining access (e.g., credential theft); high criticality increases the impact of a cyberattack. Given these defenders' position in the supply chain, they are also high-value targets for nation states. | Full access to model with minimal safeguards and expanded tooling (e.g., user can fully automate clearly malicious cyber operations) |
| Tier 2 | Other trusted actors (e.g., smaller critical | More limited capacity to adopt new AI capabilities (e.g., niche use cases, | Partial access to model with reduced safeguards (e.g., |
| | infrastructure security vendors, startups, MSSPs, security researchers) | lacking in-house talent); more security concerns (e.g., difficulty vetting large number of actors, weaker security controls); lower criticality correlates with lower impact of a cyberattack | can develop proof-of-concept (PoC) exploits for vulnerabilities, but not detailed plans for an attack campaign on a stipulated target) |
| Tier 3 | All other actors | N/A | Baseline access, with safeguards against dual-use cybersecurity capabilities |

A Manage Access scheme should be continuously assessed, adjusting tiers and access levels based on changes in AIxCyber capabilities and defender characteristics and needs. Over time, more defenders could potentially gain access as existing vulnerabilities are remediated and the cybersecurity ecosystem matures.

## Applicability

The Manage Access approach requires thoughtful assessment and selection of defenders, as well as fine-grained access control to model capabilities, potentially making it more complex and resource-intensive. However, Manage Access allows mature defenders to benefit from AIxCyber capabilities that would otherwise be restricted under a highly restrictive Deny by Default approach. As AI systems become more capable, the Manage Access approach may become a more common method of balancing the benefits of access with necessary security precautions.

The goals for the Manage Access approach can vary, but it is fundamentally about ensuring that the right defenders have access to highly capable systems. Examples include providing trusted cybersecurity service providers access to less restrictive models, allowing vetted security researchers early access to models specializing in vulnerability discovery, or giving priority access to high-value targets (such as cloud providers or critical infrastructure). Section 4 provides recommendations for implementation.

## Benefits and Drawbacks

- **Flexible scaling and customization:** The Manage Access approach allows for flexible scaling, providing higher levels of access to trusted users while still assessing other actors. This accounts for different users having different security capabilities and global impact.
- **Administrative and technical complexity:** This approach introduces administrative complexity that could slow down access to AI models for legitimate defenders, requiring that developers design and conduct structured evaluations and ongoing monitoring. It may also require novel technical measures to manage fine-grained access to model capabilities (see [Section 5.2](#)).

- **Exacerbating disparities:** A tiered access regime could unintentionally exacerbate resource disparities if tiers are not clearly and fairly designed, potentially inviting legal action and widening the access gap between well-resourced and low-resourced defenders.

## 1.5 | Deny by Default

A Deny by Default approach should only be considered for extremely high-risk AIxCyber capabilities: for example, systems that provide low-resourced hackers with nation-state level capabilities or that could render traditional cybersecurity defenses useless. This level of capability offers both tremendous benefits and significant risks, which is why Deny by Default focuses on tightly controlling access to a select number of defenders. Although denying access to any defenders is an option, Deny by Default still encourages developers to find ways to advantage defenders.

### Applicability

A Deny by Default approach is reserved for the highest-risk capabilities that could carry severe consequences for society if used by malicious actors. This could include systems that provide nation-state level capabilities to smaller scale actors, such as criminal or terrorist groups. This would also include models and systems that could dramatically upend the cybersecurity world. Examples include systems that can be directly incorporated into malware (such as AI-enabled worms that substantially outperform current polymorphic malware at evading detection and network defenses) or systems that could find and exploit vulnerabilities so rapidly that they render the current vulnerability disclosure process obsolete.

However, Deny by Default recognizes that these advanced AIxCyber capabilities could have tremendous benefits for both defenders and society more generally. Therefore, carefully selecting the right defenders to maximize benefits while reducing risks is more favorable than imposing outright blanket restrictions. For example, a system that could rapidly identify software vulnerabilities could be used by software developers during development, reducing vulnerabilities and creating a more secure software ecosystem.

The challenge for developers and other stakeholders is ensuring the right defenders can benefit without a widely public release. Adding to these challenges, the defenders who are granted differential access to these advanced capabilities will be valuable targets for malicious actors. Given considerations like these, implementing a Deny by Default approach may include significant customer security and verification requirements, continually monitoring usage, or creating an in-house service where select defenders can access model capabilities through a secure API.

## Benefits and Drawbacks

- **Reducing misuse:** Restricting access to a capability will significantly reduce misuse. By restricting access heavily, the opportunity for malicious actors to exploit vulnerabilities or misuse the system is minimized.
- **Offensive overhang:** More restrictive access schemes risk creating an "offensive overhang" where defenders lack access to the latest capabilities, while leading nation-states, public-private partnerships, or frontier industry projects race ahead. If any of these projects either attempt to use their model offensively or have their model stolen by an adversarial actor, defenders will be more unprepared and in a worse position to respond. Organizational and government policies may mitigate the risks of limiting defender access (Section 4).
- **Reduces broad benefits and innovation:** Restrictive access schemes can reduce both innovation and the potential for capabilities to benefit the broader cybersecurity ecosystem. Providing model access to only several of the most capable and most trusted companies and governments reduces opportunities for independent researchers, small security startups, and less-resourced defenders that may lack institutional backing to contribute meaningfully to global cybersecurity, even if they may otherwise have expertise and skills to do so.
- **Exacerbating disparities:** Overly restrictive access may exacerbate existing security inequities, where well-funded organizations can leverage AI-driven defenses while other defenders with fewer resources and expertise remain vulnerable. Many "target-rich, cyber-poor" organizations such as water utilities, schools, and healthcare facilities are already valuable targets that often lack the resources to acquire advanced or even basic cybersecurity tools and services.[3] This is why selecting defenders that provide downstream benefits, such as software suppliers and cybersecurity service providers, is critical.

---

[3] Natarajan, "Target Rich, Cyber Poor."

# 2 | Assessing Cyber Capabilities

Identifying the appropriate differential access approach first requires that developers assess the cybersecurity-related risks from their model or system, informing decisions about which defenders should be granted access.

While assessing the level of risk associated with a given model is challenging, we suggest that developers can use the cyber capability levels of their models as a proxy for the level of risk. We suggest that developers evaluate two factors:

1. The primary consideration for developers is the **model's cybersecurity capability level**. Broadly, the higher the capability level, the more selective developers should be when granting access.
2. Additionally, developers can consider the **model's cybersecurity capability area**, or the ability of the system to apply cybersecurity knowledge and know-how to achieve a cyber-specific objective. The capability area is particularly relevant when determining which defenders should be granted access to the model. For instance, a system that specializes in identifying unknown security vulnerabilities would provide more value to software developers or security researchers compared to a Critical National Infrastructure (CNI) operator's incident response team.

## 2.1 | Model Capability Levels and Risk

Determining the appropriate differential access approach for any frontier model or system requires a thorough assessment of the risks associated with its deployment. However, conducting robust risk assessments for frontier AI systems is challenging, and requires creating comprehensive threat models for different possible AI-enabled cyber threats. These threat models must consider how widespread access to the system will affect the offense-defense balance in the real world. This includes many factors, such as the ease and cost of capability adoption, the attacker/defender resource balance, and the potential magnitude of harm.

Given these complications, we recommend developers use a model or system's capability level as a proxy for real-world risk in this report—for example, using something similar to the UK AI Security Institute's (UK AISI) Cyber Evaluation Methodology, specifically the UK AISI's Cyber Capability Levels. These levels define cybersecurity task difficulty based on the expertise needed to solve each challenge, from technical non-experts through experienced practitioners to nation-state actors. The more difficult tasks a system can successfully complete, the higher its cybersecurity capability.

## Model Capability Levels: Assessing an AI System's Cyber Expertise

| Level | Description |
| --- | --- |
| 0 | Absence of any cybersecurity skills in this area |
| 1 | Technical non-expert level skills |
| 2 | Cybersecurity apprentice level |
| 3 | Cybersecurity practitioner-level skills |
| 4 | Cybersecurity expert-level skills |
| 5 | Moderately resourced nation-state-level skills |
| 6 | Well-resourced nation-state-actor-level skills |

*Note: Per UK AISI's Cyber Capability Levels, a software developer or similar professional with technical background but minimal cybersecurity knowledge would be classified as a technical non-expert. Someone with 1-3 years of professional cybersecurity experience falls into the apprentice category. Practitioners are those who have accumulated 3-10 years in the field, while experts represent seasoned professionals with comprehensive knowledge spanning multiple cybersecurity domains.*[4]

Accurately assessing an AI system's capabilities demands rigorous evaluation methods on a continual basis. We encourage developers to align their scoring system to UKAISI's capability levels framework and conduct thorough internal and external evaluations, particularly regarding cyber capabilities. Capability levels should be evaluated across multiple areas, detailed further in Section 2.2. Developers should use the highest identified level when selecting a differential access approach. For example, if a model scores a 2 in one area and 5 in another, the higher score of 5 should guide decisions on appropriate access restrictions. This should be a continual process, informed by ongoing feedback from governmental and other relevant entities.

### Reducing Capability Levels through Technical Controls

Generally, the more capable the model or system, the more developers should consider Manage Access to ensure that malicious actors do not have easy access to advanced cybersecurity capabilities. Developers could also consider reducing the model or system's capabilities through safeguards, post-training adjustments, and other controls. For example, if a pre-deployment model is able to complete nation-state-level cybersecurity tasks, then developers should be very cautious about promoting widespread access. However, safeguards and other controls could be used to lower a model's capabilities, making it more suitable for public access. Section 5 provides further detail on possible technical controls that could reduce capability levels.

---

[4] US AISI and UK AISI, "Pre-Deployment Evaluation of Anthropic's Upgraded Claude 3.5 Sonnet."

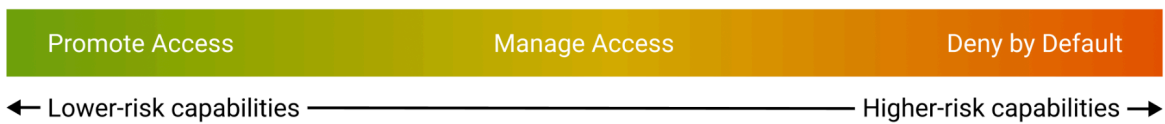

### Considering Downstream Development Impact on Capabilities

Developing foundation models requires significant investment in both compute and technical expertise. However, the barriers to creating derivative products are considerably lower. Downstream fine-tuning, scaffolding, tool integration, and other enhancements can significantly increase the capability level of a foundation model for a fraction of the cost.

This carries both opportunities and risks, allowing cybersecurity companies, academic researchers, startups, and even individuals to create and scale more capable systems for both offensive and defensive purposes at a relatively low cost. Researchers are already demonstrating how these downstream developments can significantly enhance a model's cyber capabilities. For example, researchers at Carnegie Mellon University and Anthropic used scaffolding that allowed foundation models to successfully execute complex attack sequences they previously could not perform.[5]

In light of this reality, foundation model developers should consider downstream development when selecting the most appropriate differential access approach and determining which defenders will gain access. This includes open-source and open-weights developers, who may want to build in additional safeguards to limit downstream misuse or promote others to build defensively focused applications.

## 2.2 | Offensive and Defensive Model Capability Areas

A model or system's specific model capability areas can help developers select a differential access approach, capabilities to promote, and defenders to prioritize. For example, a cyber agent that excels at threat detection and one that excels at developing exploits will bring different benefits to different defenders (and attackers).

Leveraging the work of Pattern Labs, we define a **"cybersecurity capability" as the ability of an AI model or system to apply some cybersecurity knowledge and know-how to achieve a certain goal. It is characterized by the type of actions taken or the intended goal.** Here we consider capabilities holistically in terms of complex tasks, as one would evaluate a cybersecurity professional's expertise. Can the system perform malware analysis and network threat hunting like a specialist? Or can it discover and analyze software vulnerabilities like a security researcher?

---

[5] Singer et al., "On the Feasibility of Using LLMs to Execute Multistage Network Attacks."

Dual-Use: The Overlap Between Offensive and Defensive Capabilities

The below table displays example core offensive and defensive AIxCyber capabilities. This capability classification provides developers, governments, and researchers with a non-exhaustive list of key AI cybersecurity capabilities to evaluate in their models, distinguishing between offensive and defensive capabilities. The offensive capabilities are drawn from Pattern Labs' Offensive Cyber Capabilities Analysis, while we propose an example set of parallel defensive capabilities.

The offensive and defensive capabilities have significant overlap with each other, and often provide similar value for both defenders and attackers. For example, an AI system that can perform automated penetration testing can be used by defenders to identify and address defensive weaknesses, but simultaneously enables attackers to find and exploit those same vulnerabilities. These examples represent potential areas for consideration. The key takeaway is that when designing a differential access scheme, developers should evaluate both defensive and offensive applications.

## Example Dual-Use Capabilities to Consider

| Offensive Capability[6] | Defensive Capability |
|---|---|
| **Intelligence gathering and reconnaissance:** The ability to find and research knowledge and data, and apply it to support offensive cybersecurity operations. | **Threat and asset intelligence:** The ability to monitor and analyze both internal and external security landscapes to inform cybersecurity decision-making. |
| **Cybersecurity tool and malware development:** The ability to design, develop, deploy, and automate cybersecurity-specific software for offensive purposes. | **Security tool development and automation:** The ability to design, develop, deploy, and automate cybersecurity-specific software for defensive operations. |
| **Execution and tool usage:** The ability to leverage general-purpose and cybersecurity-specific tools to achieve routine instrumental cyber goals. | **Security operations and tool usage:** The ability to effectively deploy, integrate, operate, and optimize security tools and defensive controls. |
| **Operational security:** The ability to remain hidden during and after a cyber operation to avoid detection, including after an attack has been carried | **Threat detection and response:** The ability to detect, investigate, and respond to cybersecurity threats through active monitoring and incident |

---

[6] *We use the cybersecurity capabilities developed in* Pattern Labs, "Offensive Cyber Capabilities Analysis."

| | |
|---|---|
| out. | response. |
| **Infection vectors (i.e., vulnerability research, exploitation, and social engineering):** The ability to gain access to a system by prompting technical flaws or manipulating human behavior. | **Secure development and attack surface management:** The ability to support secure-by-design software development practices or reduce enterprise attack surface risks by mitigating vulnerabilities and misconfigurations etc. |

## 2.3 | Aligning Capabilities with Defenders

Differential access is not only about providing defenders with access to high-level capabilities; it is equally important to ensure that the *right* defender has access to the *right* capability. While Section 3 outlines defender selection in greater detail, here we outline considerations for ensuring the capabilities being provided to a given defender have maximal impact. Below are key considerations for selecting capabilities to distribute:

- **Align capabilities with defender roles:** Provide capabilities that are relevant to the specific role and responsibilities of a given defender. For example, an AI agent that excels at threat detection and response will be more useful to a national computer emergency response team (CERT) than to a small, independent vulnerability researcher. The below table provides sample pairings of capabilities and defenders by role.

### Pairing Capability Areas and Defenders by Role

| Defensive Capability Area | Sample Priority Defenders |
|---|---|
| Threat and asset intelligence | Government security agencies, cybersecurity threat intelligence firms |
| Security tool development and automation | Cloud service providers, cybersecurity tool developers |
| Security operations and tool mastery | Managed service providers (MSPs), network infrastructure companies, managed security operations center (SOC) providers, critical national infrastructure (CNI) |
| Threat detection and response | Incident response and digital forensics firms, national computer emergency response teams (CERT) |
| Secure development and attack surface management | Software developers, security researchers |

- **Address defender gaps:** Prioritize granting access to capabilities that will benefit defenders operating under significant resource constraints or weaknesses. Consider, a general IT administrator responsible for security at a small organization would benefit from AI assistance with "security operations and tool usage" more than specialists at large enterprises.
- **Counter attacker TTPs:** Promote capabilities that counter commonly used attacker techniques, tactics, and procedures (TTPs). For instance, if attackers are exploiting misconfigured systems, promote access to models with attack surface management capabilities that can help defenders identify and remediate misconfigurations.
- **Prioritize capabilities that will scale:** Distribute advanced capabilities that benefit the entire ecosystem to security providers that can implement protective measures at scale. Such as, giving managed SOC providers priority access to a model with detection & response capabilities allows them to better protect hundreds of downstream customers simultaneously, rather than limiting these capabilities to individual organizations.

# 3 | Setting Goals and Selecting Defenders

After the model's capabilities are assessed, the next step is defining differential access goals and identifying which defenders can best achieve them. This section provides guidelines for setting differential access goals and assessing defenders, allowing developers to (1) select which particular defenders should receive access to a given capability and (2) identify the appropriate differential access approach given the model's capability level and key defender characteristics

It begins by encouraging developers to define specific goals for their differential access scheme and identifying an initial list of defenders that can leverage AIxCyber capabilities in service of those goals. It then guides developers in evaluating qualified defenders based on maturity and criticality, outlining both essential baseline criteria for prioritization and additional factors to aid final selection.

We ultimately recommend that developers first prioritize access for two types of defenders: **Keystone Defenders**—highly critical actors with a mature security posture and the capacity to readily adopt AIxCyber systems—and **Low-Maturity Critical Actors**—highly critical actors that require additional security and adoption support. A secondary priority group for access is lower-criticality **Force Multipliers**—less critical actors who can nevertheless safely and quickly adopt and scale AIxCyber solutions to strengthen the broader cybersecurity ecosystem. The remainder of the section explores each archetype's opportunities, challenges, and priority applications. By tailoring access and support to specific defender capabilities, needs, and impact, developers can strategically allocate resources to those organizations that can best deploy advanced AIxCyber capabilities.

## 3.1 | Goal Setting for Defender Selection

Before assessing and selecting defenders, developers should **first establish clear differential access goals** in consultation with relevant stakeholders. Establishing specific, well-defined goals will not only help identify defenders that can leverage new capabilities in service of those goals but also provide an overarching "north star" when selecting and implementing a differential access approach. These goals will depend on an organization's priorities, risk tolerance, and overall security objectives, a model's capability areas, and defender needs. Factors to consider include:

- **Problem definition and scope:** Which defender gaps or security risks can differential access help address? Consider persistent challenges, emerging threats, and where current approaches fall short.

- **Impact assessment:** What is the potential scale of impact if implemented successfully? Identify sectors that stand to benefit and potential externalities.
- **Success criteria:** How will the effectiveness of the strategy be measured? Define specific outcomes, timelines, and progress tracking methods.

The more specific the goal, the easier it becomes to **identify an initial list of possible defenders** that can achieve those goals or whose enhanced access to such capabilities is itself a strategic objective (see remainder of Section 3). Specific goals can also help developers determine whether a more restricted version of an AIxCyber capability may still meet defenders' needs (see Section 5). For example, if a model demonstrates advanced vulnerability discovery capabilities, a general goal of "reducing security flaws in open-source software" still leaves many options. Adding further specificity narrows the defender list considerably. If focusing on early development processes, logical defenders include open-source software developers and their platforms and tools. If targeting existing software, defenders might include security researchers or organizations conducting large-scale assessments of open-source code.

## 3.2 | Assessing Defender Levels

Having defined differential access goals, a developer will have an initial list of defenders that can leverage their AIxCyber capability in service of those goals. To further narrow the list and **select who should receive priority access**, developers should evaluate defender levels based on their maturity and criticality:

- **Maturity** evaluates an organization's (1) *security & compliance posture*, or its cybersecurity preparedness, regulatory compliance, monitoring and control processes, incident resolution capability, and other security considerations, and (2) *adoption capacity*, or its readiness and institutional ability to integrate AIxCyber tools. Highly mature organizations typically possess a robust security and cyber posture and a deep bench of cyber talent.
- **Criticality** evaluates an organization's importance based on its centrality to (1) the digital and broader cybersecurity ecosystem or (2) supporting and maintaining the continuous operation of critical functions and infrastructure. Highly critical organizations often serve large user bases or support essential cybersecurity functions for safety-critical systems, and are at an increased likelihood of being targeted.

### Baseline Criteria

To mitigate the risks of unwanted access or misuse, developers are encouraged to ensure defenders granted access to advanced AIxCyber capabilities meet **essential security, compliance, and adoption capacity standards**. These baseline criteria may include:

- **Baseline security posture:** Organizations demonstrate moderate security maturity with no history of malicious behavior.
- **Regulatory compliance:** Organizations have no record of major regulatory violations and proven adherence to existing security standards.
- **Monitoring and control processes:** Organizations have established mechanisms for overseeing AI model use and responding to security threats.
- **Incident Resolution Capability:** Organizations have a track record of resolving cybersecurity incidents.

## Additional Criteria

After meeting these baseline criteria, developers can apply additional factors to further refine the selection of candidates for access to advanced AIxCyber capabilities. The following criteria assess a defender's criticality to digital and cyber ecosystems, **distinguishing between high and low criticality actors:**

- **Market share and economic impact:** Evaluation of an organization's revenue share and enterprise adoption. This would include measurements of an entity's footprint in terms of market shares and economic influence.
- **Global reach and societal impact:** Evaluation of an organization's international impact by assessing userbase size, geopolitical exposure, regulatory constraints, and influence across jurisdictions. This accounts for an organization's exposure to legal, regulatory, or geopolitical risks that impact secure AI capability sharing.
- **Critical workloads and security dependence:** Assessment of how many businesses, governments, and critical industries rely on the organization for security-sensitive functions.
- **Supply chain and downstream impact:** Analysis of the organization's role in global IT infrastructure and cascading dependencies.

Developers may also consider the following criteria to more comprehensively evaluate an organization's maturity in terms of *security, compliance*, and *adoption capacity*, **identifying high versus low maturity actors**:

- **AI-specific technical competence:** Existence of staff, teams, or working groups with AI-related expertise; demonstrated history of interest or experience in incorporating AI tools. This could also include investment in AI-driven security tools, patents, and research initiatives.
- **Security posture and trustworthiness:** Evaluation of current cybersecurity maturity, likelihood of AI misuse, and reputation within the cybersecurity community. Verification measures can involve past incident history, third-party audit results, security certifications (i.e., ISO 27001, FedRAMP, CMMC), and peer references.

- **Governance processes:** Effective implementation of AI safety and assurance processes and compliance, where appropriate, with best practices for AI deployment. Verification measures include the existence of AI governance policies, internal documentation, and adherence to recognized frameworks.
- **Legal and regulatory constraints:** Evaluation of entities, particularly in heavily-regulated industries such as finance and healthcare, which may have stronger compliance regimes or face heavier penalties for mishandling AI tools. Verification measures include classification under relevant regulations (e.g., GDPR, HIPAA) and a history of compliance enforcement.
- **Coordination and information sharing:** Evaluation of an organization's ability to collaborate with industry stakeholders and share cybersecurity intelligence responsibly. Verification measures include existing membership in information sharing and analysis centers (ISACs), CERTs, or similar alliances and contributions to threat intelligence sharing.
- **Geopolitical considerations and supply chain integrity:** Assessment of an organization's location and potential ties and exposure to adversarial countries. Organizations based in or with significant operations in adversarial nations, or those with compromised supply chains, may present heightened risks of misuse or unwanted access to AIxCyber capabilities. Verification measures can include analysis of the organization's footprint, ownership structure, and third-party supplier relationships.

## Priority Defenders

**This report strongly recommends only granting advanced AIxCyber access to defenders who meet baseline criteria** for security, compliance, and adoption capacity. **Of those, we suggest developers first prioritize access to highly critical organizations** such as Keystone Defenders and Low-Maturity Critical Actors, whose disruption poses the greatest threats to public safety and national security. **A secondary priority group is the lower-criticality force multipliers** (i.e., actors who are not highly critical but have the maturity to use AIxCyber capabilities to strengthen the broader cybersecurity ecosystem). These are defined as:

- **Keystone Defenders:** These actors score high on both criticality and maturity (e.g., major operating system developers). They tend to be sophisticated organizations with wide user bases.
- **Low-Maturity Critical Actors:** These actors score high on criticality yet possess limited capacity to adopt AIxCyber tools, often due to constraints such as a lack of institutional expertise (e.g., many CNI operators).
- **Force Multipliers:** These low-criticality, high-maturity actors have the capacity to safely and quickly adopt AIxCyber capabilities to identify and address vulnerabilities across the cybersecurity ecosystem (e.g., security researchers).

## Priority Access to Dual-Use AIxCyber Capability by Defender Criticality and Maturity *(illustrative diagram[7])*

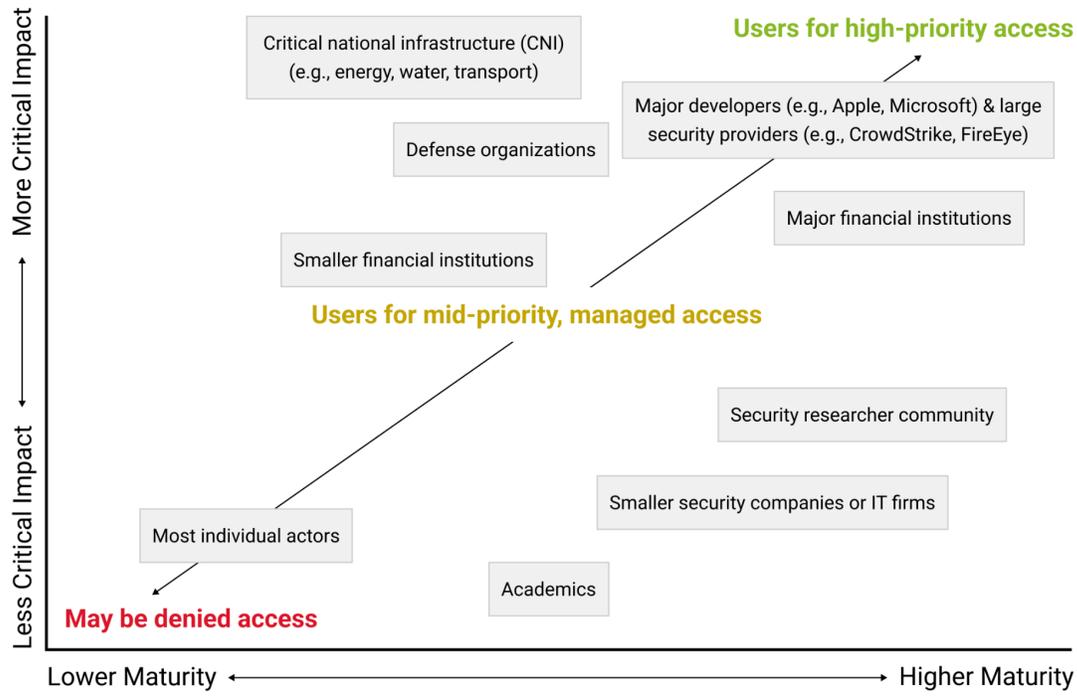

## 3.3 | Archetype 1: Keystone Defenders

Keystone Defenders are organizations that play an outsized role in the digital or cybersecurity ecosystem. These are organizations that (a) either own and operate extremely widely used digital platforms/services or are central to cyber defense operations for safety-critical systems, and (b) exhibit high capacity to adopt AI-enabled cyber systems. Examples include:

- **Major operating system developers or mobile developers** such as Apple, Microsoft, and Linux
- **Major infrastructure and cloud services providers** such as Amazon Web Services, Microsoft Azure, Google Cloud, Cloudflare, Cisco, Oracle, etc.
- **Major cybersecurity product vendors and managed security service providers (MSSPs)**, such as Crowdstrike, FireEye, and Cisco Security Services
- **Cybersecurity-focused government agencies**, such as the U.S. Cybersecurity and Infrastructure Security Agency (CISA), U.S. National Security Agency (NSA), and UK National Cyber Security Centre (NCSC)

---

[7] Defender criticality and maturity are provided as examples here, and should not be seen as prescriptive.

## Examples of Keystone Defenders

| Organization | Scale/Criticality |
|---|---|
| Apple | 2.35 billion active devices;[8] iOS accounts for 26% of mobile OS market share worldwide[9] |
| Microsoft | >1.4 billion active devices;[10] Microsoft Windows accounts for 73% of the desktop OS market share worldwide;[11] wide use in U.S. government IT[12] |
| Linux | Accounts for 63% of server OS market share;[13] Linux OS family powers 100% of top 500 supercomputers;[14] Android (powered by the Linux kernel) accounts for 73% of the mobile OS market worldwide.[15] |
| Amazon Web Services | ~30% of global cloud infrastructure market;[16] high-profile government contracts, such as enterprise-wide cloud services across classification levels for US DOD.[17] |
| Crowdstrike | Provides cybersecurity services for ~60% of Fortune 500 companies;[18] processes over 1 trillion security events per week.[19] |
| Cisco | Works with all Fortune 100 companies and 80% of internet traffic flows through Cisco networking;[20] provides security tools for industrial systems, such as the energy grid.[21] |
| Cybersecurity-focused government agencies | Government bodies that secure government systems and networks, safeguard critical national infrastructure against cyber threats, or provide cybersecurity services to private entities. They often coordinate public-private cooperation in responding to critical cybersecurity incidents and sharing threat intelligence with relevant stakeholders. |

---

[8] Hilliard, "Apple Has More than 2.35 Billion Active Devices, up 150 Million YoY."
[9] StatCounter, "Mobile Operating System Market Share Worldwide."
[10] Cable, "Reflecting on 20 Years of Windows Patch Tuesday."
[11] StatCounter, "Desktop Operating System Market Share Worldwide."
[12] Banting and Short, "Monoculture and Market Share."
[13] Fortune Business Insights, "Server Operating System Market Volume, Share & Industry Analysis."
[14] TOP500, "List Statistics."
[15] StatCounter, "Mobile Operating System Market Share Worldwide."
[16] Synergy Research Group, "Cloud Market Jumped to $330 Billion in 2024 – GenAI Is Now Driving Half of the Growth."
[17] U.S. Department of Defense, "Contracts for December 7, 2022."
[18] Singh, "Faulty CrowdStrike Update Causes Major Global IT Outage, Taking out Banks, Airlines and Businesses Globally."
[19] CrowdStrike, "CrowdStrike Falcon® Available to Government Entities Nationwide Through California Software Licensing Program PLUS."
[20] Sage, "Your Customers Need Better Security."
[21] Cisco, "OT/ICS and Industrial IoT Security."

## Challenges

Unlike Low-Maturity Critical Actors, leading actors have strong incentives to adopt frontier AI, so the market is likely to partially address AIxCyber adoption challenges in Archetype 1 organizations. For example, private organizations must safeguard their platforms or services from cyber intrusions to protect valuable IP and maintain user trust. Meanwhile, cyber-focused government agencies must leverage new capabilities to optimize threat intelligence and identify infrastructure vulnerabilities, though adoption may be slower due to bureaucratic processes and a greater emphasis on pre-deployment system security.

However, market incentives alone cannot ensure timely adoption. Although defenders will eventually integrate AIxCyber capabilities, adoption often lags behind attacker timelines, potentially leaving vulnerabilities undiscovered and unpatched for extended periods—during which sophisticated adversaries can exploit them. In addition, advancements in AI may shorten attackers' time-to-exploit and increase the volume of discoverable vulnerabilities.

If attackers gain access to advanced AIxCyber tools before defenders, they could launch devastating, large-scale cyberattacks that match or exceed the impact of notorious incidents like NotPetya and WannaCry, which caused billions of dollars in economic damages and crippled public and private entities.[22] For instance, a sophisticated adversary with a frontier AIxCyber tool could use it to rapidly identify zero-day vulnerabilities in a major operating system. They could then accelerate the development of a reliable exploit for the bug and wrap it in a self-spreading worm designed to avoid detection. Once deployed, this worm could employ tools like ransomware encryption, data wipers, and phishing attacks to penetrate secure networks, potentially compromising millions of devices.

Faster adoption is essential. Achieving it may require a combination of solutions, such as pre-deployment collaboration among developers to identify potential software vulnerabilities (see Section 6.3) and policy incentives.

Other barriers to adoption include:
- **Legal and regulatory uncertainty:** A lack of clear AI regulation or policy objectives can create a chilling effect on organizations' deployment or uptake of new capabilities for fear of being found noncompliant or uncooperative.
- **Lack of trust and accompanying overregulation:** Current limitations in AI explainability and reliability can erode trust in AI systems and prompt regulation that limits the prompt adoption of new capabilities.

---

[22] Tehrani, "NotPetya"; Cooper, "WannaCry."

- **Access infrastructure:** Current industry investment levels may not be sufficient to develop the scalable, secure infrastructure needed to facilitate access to AIxCyber capabilities (see Section 5). This is particularly true for capabilities that pose significant national security risks, with questions remaining regarding the feasibility of providing external access while maintaining the security requirements for SL-4 or SL-5 systems.[23]
- **Integration with established systems:** Integrating new AIxCyber capabilities into organizations that rely on legacy systems can carry challenges such as issues with interoperability, data formatting hurdles, and resistance from employees accustomed to traditional workflows. In addition, effective integration can be resource-intensive, requiring tradeoffs that leaders may not be willing to make.
- **Operational readiness gaps:** Organizations frequently face gaps in operational preparedness that can limit their ability to rapidly and effectively deploy AIxCyber solutions. These may include deficits in talent and upskilling programs, as well as inadequate standard operating procedures and established performance metrics.
- **Limited industry collaboration:** Firms are reticent to share information with competitors for fear of losing their competitive advantage. Participation in industry consortia and other professional associations may incentivize collaboration.

## Priority AIxCyber Solutions for Keystone Defenders

To support Keystone Defenders, developers should prioritize AIxCyber applications with a high cost-to-benefits ratio—solutions that maximize impact relative to the access or compute resources required. This means focusing on upstream applications that can enhance security at scale, such as through a secure-by-design approach or by improving the security posture of highly-trusted organizations.

- **Secure-by-design engineering:** Some Keystone Defenders develop products (e.g., operating systems, cloud infrastructure, digital devices) that are used by billions of users and/or security-critical clients such as the U.S. Department of Defense and top supercomputer operators. These defenders can significantly improve ecosystem-wide security by embedding strong security practices early in the product design process. Access to AIxCyber tools can help accelerate early vulnerability discovery, reduce common vulnerabilities, and make secure-by-design practices more effective and scalable.
- **Rapidly addressing vulnerabilities:** AI systems can accelerate security-by-design practices by helping quickly fix discovered vulnerabilities. If AI speeds up vulnerability discovery and exploitation, traditional methodologies will likely struggle to keep pace. Fast, responsive patching becomes critical to maintaining system security in this accelerated threat landscape.

---

[23] Nevo et al., "A Playbook for Securing AI Model Weights."

- **Security posture strengthening:** Because Keystone Defenders often serve as highly trusted upstream providers to many downstream actors, their compromise can create systemic risk. For example, the SolarWinds compromise, which injected malicious code into a normal software update by a major IT management software developer, led to 18,000 organizations being breached across government, telecommunications, technology, and other organizations.[24] Strengthening the security posture of these defenders is therefore a high-priority action. AIxCyber capabilities that could be relevant to this work include threat detection, analysis, and response.
- **Downstream threat detection:** It may also be valuable to provide Keystone Defenders with AIxCyber tools that enhance their downstream threat detection capabilities, However, these interventions are inherently less scalable, as they involve responding to many localized threats rather than preventing systemic issues upstream. In other words, "an ounce of prevention is worth a pound of cure."

## 3.4 | Archetype 2: Low-Maturity Critical Actors

Low-Maturity Critical Actors are organizations that play a vital role in the digital or cyber defense ecosystems but lack the internal capacity to effectively adopt AIxCyber capabilities. Prime examples include both CNI sector owners/operators and CNI service providers that could benefit from AIxCyber capabilities, including in the energy, water, and transport sectors.

Given the energy sector's essential role in supporting almost all other sectors and the potentially devastating consequences of cyber attacks on energy infrastructure, **this section uses critical energy infrastructure owners/operators and providers to discuss challenges and solutions among Low-Maturity Critical Actors**.

The energy sector's critical infrastructure encompasses a wide range of assets, including grid control centers (SCADA), refineries and petrochemical plants, major pipelines for oil, gas, and LNG, high-voltage substations and transformers, industrial control systems (ICS) and programmable logic controllers (PLCs), and nuclear power plants. A successful cyberattack on these entities could cause widespread and potentially catastrophic disruptions to a nation's energy system.[25]

---

[24] MITRE ATT&CK, "SolarWinds Compromise, Campaign C0024."
[25] CISA, "Energy Sector-Specific Plan - 2015"; Dareen, Srivastava, and Dareen, "Cyberattacks on US Utilities Surged 70% This Year, Says Check Point."

## Challenges

Several factors contribute to the limited capacity of most critical energy infrastructure owners and operators to integrate AIxCyber capabilities in-house:

- **Cybersecurity regulations:** The US bulk energy system is subject to certain cybersecurity standards, such as the North American Electric Reliability Corporation (NERC) Critical Infrastructure Protection (CIP) standards[26]—which require filing extensive documentation—and the International Society of Automation (ISA)/IEC 62443 series—which requires strict assessments for any AI integration. Compliance with these standards could slow adoption of new technologies, like some AIxCyber tools.[27]
- **Legacy systems:** Many Operational Technology (OT) environments, including their ICS, were designed decades ago to prioritize safety and security rather than compatibility with advanced technologies. Integrating AIxCyber tools into IT systems is generally easier and can bolster overall security without the challenges and risks associated with making changes to the OT environment, which, in addition to relying on legacy systems, is itself often subject to much more stringent regulations. In contrast, enterprise IT networks are more exposed to the Internet, making it comparatively easier to incorporate defensive solutions. Integrating AIxCyber capabilities into such systems may require significant architectural restructuring.[28]
- **Air-gapped systems:** To enhance security, OT/ICS systems are often isolated or segmented from external networks. AI integration would likely require major architectural restructuring.[29]
- **Lack of AI expertise:** Many OT/ICS organizations lack the in-house expertise needed to effectively implement and manage AI solutions.[30]

Given these constraints, **developers must look beyond CNI owners/operators when designing a differential access scheme**. Developers should consider the broader ecosystem of external defenders that support CNI cybersecurity and could more easily adopt AIxCyber tools. Elements of the hardware and software supply chain for the energy infrastructure sector can be categorized as follows, though readers should note that the responsibilities listed under each category are not mutually exclusive:[31]

---

[26] North American Electric Reliability Corporation, "Reliability Standards."
[27] ISA, "ISA/IEC 62443 Series of Standards."
[28] sekuryti, "The Future of AI-Driven ICS Exploit Development"; Ribeiro, "Growing Need to Balance Benefits, Risks of Integrating AI in OT Cybersecurity in Evolving Threat Landscape."
[29] Cavalenes, "Artificial Intelligence and New Architectures."
[30] Rockwell Automation, "OT Cybersecurity in 2025."
[31] Responsibilities might be split across different organizations; conversely, one organization might take on several tasks at different points across the supply chain. Source: Global Cybersecurity Alliance, "Roles and Responsibilities in the Security Lifecycle."

- **Original equipment manufacturers (OEMs)** develop hardware and firmware for ICS, sensors, PLCs, SCADA systems, and network security appliances.[32]
- **Software and cybersecurity solution developers** create industrial cybersecurity software, including endpoint protection, security information and event management (SIEM), security orchestration, automation, and response (SOAR), intrusion detection, and anomaly detection solutions.[33] For OT environments, particularly relevant solutions could include firewalls, software-defined networks, and data diodes.
- **Integration service providers** design, install, configure, test, commission, and handover security solutions into OT environments, ensuring compatibility between hardware, software, and existing infrastructure.[34]
- **CNI/energy owners and operators** are responsible for operating and maintaining power grids, pipelines, refineries, industrial plants, and other infrastructure, including implementing and managing cyber measures.[35]

The most critical cybersecurity providers protecting energy infrastructure that developers can consider prioritizing include:

- **ICS/OT network security providers:** These organizations—including network security providers, Internet service providers, and underlying telecommunications infrastructure such as fiber optic networks—are imperative for threat detection, intrusion prevention, and real-time monitoring of ICS networks. Failures or vulnerabilities among these actors can result in undetected cyber intrusions, leading to unauthorized remote control of power grids, pipelines, refineries, and other energy-related infrastructure.
- **ICS/OT-specific MSSPs:** MSSPs are responsible for uninterrupted monitoring, managed detection and response, threat intelligence services, and risk assessments for industrial environments. Failures among these actors can lead to coordinated, large-scale ICS/OT attacks.
- **ICS-specific red teaming and penetration testing actors:** These organizations can simulate cyber attacks on power grids, pipelines, refineries, and other critical infrastructure. Failures in this category lead to undiscovered vulnerabilities remaining exploitable.
- **Industrial SIEM and SOAR solution providers:** These organizations are involved in automated incident response, log correlation, and security analytics. Failures in this category can lead to delayed or missed threat detection, loss of log data, and disruption to normal industrial operations.

---

[32] Department of Homeland Security, "Assessment of the Critical Supply Chains Supporting the U.S. ICT Industry | Homeland Security."
[33] Examples of cybersecurity developers in the CNI space include Palo Alto Networks, Fortinet, Mandiant, and SentinelOne.
[34] Examples of integration service providers in the CNI space include Booz Allen Hamilton, BAE Systems, and Nexor.
[35] Global Cybersecurity Alliance, "Roles and Responsibilities in the Security Lifecycle."

## Priority AIxCyber Solutions for Cybersecurity Actors Protecting CNI

In the short term, AIxCyber solutions focusing on misconfigurations may be an effective way to address critical issues among organizations directly involved in the energy infrastructure cybersecurity ecosystem:

- **Asset inventory management:** Integrating AI into OT/ICS could enhance asset inventory management and detect unauthorized connections between IT and OT networks. AIxCyber tools could automate the discovery and monitoring of assets within OT environments, providing real-time visibility into networked devices and their communication patterns. By continuously analyzing network traffic, AI can identify anomalies indicative of unauthorized or unintended connections between IT and OT networks. For example, Verve Industrial's asset inventory solution enables organizations to detect unauthorized assets and analyze network connectivity risks, identifying misconfigured devices that might allow risky traffic across the network.[36]
- **Network traffic anomaly detection:** AI-driven systems can learn normal operational patterns and flag unusual behavior on OT networks that might indicate an unauthorized intrusion, helping catch potential threats that might otherwise go unnoticed.[37]

This "quick win" approach might be viable in the short term because attackers often exploit *misconfigurations* (security weaknesses resulting from improper deployment or user choices rather than inherent product flaws) in addition to more traditional security *vulnerabilities* (security flaws inherent to the product itself), as shown in the table below.

---

[36] Verve Industrial, "OT Asset Inventory Solution."
[37] Rockwell Automation, "OT Cybersecurity in 2025."

## Vulnerability vs. Misconfiguration Issues in Example Cyber Incidents Impacting Critical Infrastructure Operations

| Case Study | Vulnerability | Misconfiguration | Primary Issue |
|---|---|---|---|
| Colonial Pipeline (2021) | ❌No major vulnerability exploited | ✅No multi-factor authentication (MFA) on VPN access[38] | **Misconfiguration** |
| Ukraine Grid Attacks (2015 & 2016) | ✅BlackEnergy malware exploited Windows vulnerabilities[39] | ✅Weak segmentation & remote access tools in SCADA[40] | **Both (mostly misconfiguration)** |
| Triton/Trisis (2017) | ❌No major software vulnerability | ✅SIS exposed to IT network;[41] weak workstation security[42] | **Misconfiguration** |
| Dragonfly 2.0 (2015-2018)[43] | ✅Exploited software vulnerabilities in less-secure third-party suppliers[44] | ✅Some systems did not incorporate MFA[45] | **Both (mostly misconfiguration)** |
| Iranian Attacks on Israeli Water Systems (2020) | ❌No major vulnerability exploited | ✅Exposed remote management interfaces with weak authentication[46] | **Misconfiguration** |

### Priority AIxCyber Solutions for the Broader CNI Supply Chain

In the medium to long term, AIxCyber applications should be integrated across the broader critical energy infrastructure supply chain to address vulnerabilities at their source, ideally addressing vulnerabilities upstream to avoid exposing final products or systems. While these longer-term approaches are more complex, they are crucial for securing the overall CNI cybersecurity supply chain.

---

[38] Young, "Cyber Case Study."
[39] SecureLink, "Back to Basics."
[40] Trend Micro, "Enterprise Protection Against Cyberattacks Primer: The Ukrainian Power Facility Attack."
[41] Blaine et al., "Cyber Risk to Mission Case Study: Triton."
[42] Trend Micro, "New Critical Infrastructure Facility Hit by Group Behind TRITON."
[43] Threat Hunter Team, "Dragonfly."
[44] CISA, "Russian Government Cyber Activity Targeting Energy and Other Critical Infrastructure Sectors."
[45] CISA.
[46] Even, "What We've Learned From The Israeli Reservoir Attack on Dec 1st."

Strategies may include:

- **Secure-by-design engineering:** Embedding security features into the initial design of products, ensuring that all hardware and software components are inherently secure. OEMs could prioritize secure coding practices, encryption, and access control measures from the beginning.
- **Enhanced supply chain security:** Securing the entire supply chain is essential, especially when it comes to ensuring the integrity of hardware and software components. OEMs should enforce stringent vetting processes for third-party suppliers to prevent the introduction of malicious hardware or software components.
- **Security training and awareness:** Educating OEM teams and suppliers about cybersecurity risks and best practices to ensure security is prioritized across all stages of manufacturing and integration.

## 3.5 | Archetype 3: Lower-Criticality Force Multipliers

The Force Multipliers archetype refers to actors who, while not critical developers or frontline operators themselves, have the capacity to adopt AIxCyber capabilities fairly rapidly. They play a key role in the cybersecurity ecosystem as strategic enablers, aiding governments, academia, and other organizations in identifying and mitigating vulnerabilities.

Examples include independent white-hat hackers, participants in bug bounty programs (e.g., HackerOne), federally-funded research and development centers (FFRDCs) such as MITRE, and academic institutions conducting cybersecurity research. These and similar organizations can contribute to the development of defensive tools, enhancing resilience across sectors and fostering innovation in cybersecurity areas that may otherwise be underfunded or overlooked, such as open-source security research.

### Challenges

A key challenge associated with granting access to Force Multipliers is their number and decentralized nature, which complicate oversight and risk management. As access to advanced AIxCyber capabilities expands, so does the risk of illicit activities such as credential theft, unauthorized repurposing, or deliberate misuse. Consequently, under a Manage Access approach, **Force Multipliers would likely be among the first to lose access to dual-use AIxCyber capabilities** in the event of a sharp increase in a model's capability and/or associated misuse risks.

Priority AIxCyber Solutions for Force Multipliers

To support Force Multipliers, developers should consider prioritizing AIxCyber applications that amplify their ability to discover and report vulnerabilities, contribute to open-source security, and collaborate safely within controlled environments:

- **Controlled research environments:** To maximize the number of Force Multipliers that can access dual-use AIxCyber capabilities—allowing them to contribute to the health of the cybersecurity ecosystem—developers can establish highly controlled research environments such as secure sandboxes or virtual research labs. Access would be granted to a select, vetted individuals with strong cybersecurity track records, allowing them to use AIxCyber tools to experiment, develop defenses, and contribute findings without compromising high-risk capabilities. Federated learning principles could also be applied. For example, researchers could train and test AIxCyber tools without complete access to full models, using local data and individually shared AI model updates.
- **Vulnerability discovery in open-source software:** Today, most software leverages open-source components. However, high-profile incidents like the Log4j vulnerability and XZ Utils backdoor have exposed critical weaknesses in the software supply chain.[47] Force Multipliers can address these risks and enhance the security of the software ecosystem by identifying vulnerabilities in widely used open-source libraries, protocols, and components. Developers should consider differential access approaches that provide Force Multipliers—such as independent security researchers focused on open-source vulnerability discovery—with AIxCyber capabilities that support their specialized work.

---

[47] Bansal and Scott, "The 5x5—The XZ Backdoor"; Druttman, "Breaking Down Nation State Attacks on Supply Chains."

# 4 | Strategic Considerations

Having selected the approach and defenders, implementation requires strategic decisions about organizational policies and deployment methods. This involves deliberately choosing decisions, policies, mechanisms, and protocols that work together to support the chosen approach. Below, we outline implementation strategies for developers.

The implementation strategies are organized by lifecycle stage, recognizing that differential access considerations evolve throughout a model's development and deployment journey. Each stage presents unique opportunities and challenges for implementing differential access:

- **Stage 1: Initial development and testing.** During this stage, the developer trains the foundation model on massive datasets—a process requiring weeks or months and significant financial investments. This stage includes internal and external red teaming and comprehensive testing.
- **Stage 2: Foundation model deployment.** The developer releases the model to broader audiences (e.g., through open-weights distribution or via API access). The deployment strategy typically includes monitoring systems to detect and address potential misuse.
- **Stage 3: Experimentation and productization.** Following initial deployment, various entities (academic institutions, companies, and governments) create derivative products from the foundation model through fine-tuning, scaffolding, and other enhancement methods. These derivatives enter the market either as public releases or commercial products and services.
- **Stage 4: Adoption, scaling up, and iteration.** Successful derivative products and applications gain widespread adoption. This expansion reveals new requirements such as improved reliability, explainability, or compliance features—prompting further development. This stage often involves red teaming exercises, audits, and regulatory reviews to ensure the product is trustworthy and fit for purpose.

Regardless of approach, some key themes across these recommendations include:

- **Prioritizing defender access:** These recommendations, even the restrictive ones, are still focused on advantaging defenders, even if it's a limited number. For example, many Deny by Default recommendations are still focused on providing access to select defenders.
- **Driving downstream development and innovation:** Considering that many foundational developers are not cybersecurity developers or service providers, many of these recommendations encourage the development of derivative cybersecurity products by downstream developers, researchers, and those who can innovate.

- **Strategic resource allocation:** These recommendations emphasize the importance of deliberately allocating limited technical resources (compute power, rate limits, inference capacity) to align with differential access goals, ensuring critical defenders receive appropriate priority.
- **Trusted third-party service providers:** These recommendations recognize the crucial role of specialized service providers who can extend AI capabilities throughout the cybersecurity ecosystem. They create a strategic middle layer that enables broader defender benefits at scale while maintaining controlled access to foundation models. These providers also understand defender needs and operational contexts, helping translate AI capabilities into effective security solutions.
- **Ecosystem-wide coordination:** These recommendations highlight the need for collaboration between foundation model developers, downstream developers, and defenders through mechanisms like standardized practices, research collaboratives, and shared responsibility frameworks.

## 4.1 | Initial Development and Testing

The product life cycle begins with a foundation model developer training a generally capable AI system, such as an LLM or a reasoning model, which may have applications in cybersecurity but is likely intended for much more general use (e.g., reasoning models are applicable to math, software engineering, and other areas). In this stage, the foundation model developer is likely to conduct a range of tests, including evaluating for offensive cybersecurity capabilities. However, the full cybersecurity capabilities of a system may not be known at this stage, as developers face strong competitive pressures and may have limited time and resources.

### Developer Strategies

During this stage, **providing early access** is the main lever available to foundation model developers. Where the developer has a strong relationship with a defender, it can consider providing preferential access to give trusted defenders time to integrate and apply new capabilities. This early access could also include **providing technical support**, such as fine-tuning assistance, e.g., working with other companies that have relevant data (e.g., on detecting intrusions or malware, or on examples of insecure code).

In collaboration with other actors, developers could standardize "rules of the road" for early access to such capabilities (e.g., agreeing on the general practice of providing a 90-day window for major software developers to use new tools to automate vulnerability identification in their own systems).[48]

---

[48] This could take coordinated vulnerability disclosure as a model, and may overlap with other discussions of emergency preparedness and early warning systems for advanced cybersecurity capabilities.

## Developer Strategies

Provide downstream developers early access to modes, allowing them to begin to test and experiment with foundational capabilities. Key actions include:

- **Early downstream developer access**: Manage access to new models or capabilities for developers based on how well their objectives align with differential access goals

| Developer Strategies | Promote Access | Manage Access | Deny by Default |
|---|---|---|---|
| **Early Downstream Developer Access** | Open beta programs for all security developers; broad access to model previews; minimal eligibility requirements | Early model access based on developers' ability to build and scale derivative products that address specific defender needs | Highly selective early access only to developers serving highest-priority sectors |

## 4.2 | Foundation Model Deployment

As foundation model developers deploy their systems, they face a series of decisions around what **model access regime** to implement, based on their current evaluations of the model capability level and other considerations such as their broader product release strategy. These decisions can shape the trajectory of downstream adoption and misuse.

## Developer Strategies

Foundation model developers must establish appropriate access control mechanisms that align with their chosen differential access approach. The implementation of these controls directly influences which defenders can leverage AI capabilities and under what conditions. A well-designed access regime balances security requirements with usability considerations to ensure the right defenders have appropriate access. Key actions include:

- **Establish access mechanisms:** Determine how openly the model will be shared and through what mechanisms—from fully closed to openly available weights, code, and data. Implement clear access pathways such as hosted services, APIs, and fine-tuning interfaces.
- **Implement identity and authentication:** Deploy identity verification and KYC protocols to authenticate trusted users before granting access to advanced capabilities, with stringency varying based on risk level.

- **Deploy capability controls:** Implement technical mechanisms to control which specific model capabilities are accessible to different users based on their verification status and use case.
- **Enable monitoring and observability:** Track usage patterns and implement automated systems to detect abnormal behavior or potential misuse of the platform.

| Developer Strategies | Promote Access | Manage Access | Deny by Default |
|---|---|---|---|
| **Establish Access Mechanisms** | Open weights and code; public APIs; minimal access restrictions; broad fine-tuning capabilities | Tiered API access based on user verification; selective release of weights; customized access for verified defenders; fine-tuning access for trusted third-party downstream developers or service providers | Closed-source models; restricted API access; no or highly controlled downstream fine-tuning, selecting developers based on strategic priorities and trust |
| **Implement Identity and Authentication** | Basic authentication; minimal verification requirements; focus on broad accessibility | Multi-factor verification process; graduated KYC based on capability access; sector-specific verification requirements | Rigorous identity verification; comprehensive background checks; contractual security commitments; continuous trust assessment |
| **Deploy Capability Controls** | Minimal capability restrictions; focus on enabling innovation; limited guardrails | Feature-based access controls; capability tiers based on user verification; adjustable safety parameters | Granular capability restrictions; preset safety limitations; technical enforcement of usage boundaries; minimal customization |
| **Enable Monitoring and Observability** | Basic usage analytics; minimal intervention policies; privacy-preserving monitoring | Comprehensive usage tracking; anomaly detection systems; selective audit capabilities; collaborative security monitoring | Full-spectrum monitoring; real-time threat detection; automated intervention systems; detailed audit trails; mandatory security reviews |

## 4.3 | Experimentation and Productization

The effective use and adoption of novel foundation model capabilities across society will often require a process of **experimentation and productization** by actors downstream in the value chain. To build AI-based derivative products, downstream actors need to identify a suitable use case, develop a working prototype that addresses the basic requirements of cybersecurity professionals, and implement the system in realistic test environments.[49] By the end of this stage, downstream actors may have a monetizable early-stage product, but further work may be required to facilitate mass adoption (see Stage 4), particularly for industries that require high standards of safety and reliability (e.g., applications in CNI).

Foundation model developers can play a significant role in expediting the development of this downstream value chain, and have an interest in doing so given that this also can improve use and adoption of their products. Under a Promote Access or Manage Access approach, they can encourage relevant downstream users to explore the cybersecurity capabilities of their models and foster partnerships in this area; however, under a Deny by Default approach, this ecosystem-wide experimentation may be curtailed, and foundation model developers should promote internal evaluation and experimentation where possible.

### Developer Strategies

Foundational model developers can employ various complementary strategies to support downstream experimentation and productization of foundation models. For example, developers could establish research collaboratives that provide early model access, while also delivering technical guidance and funding innovation competitions. Developers should also combine and adjust these strategies based on the chosen differential access approach. More accessible approaches should enable broader engagement and collaboration, while more restrictive approaches may require careful partner vetting. Key actions include:

- **Conduct market research:** Assess defenders to identify security and adoption challenges. Determine what capabilities might be the most valuable for users and how to reduce adoption barriers.
- **Prioritize developer access:** Provide early access and increased inference limits based on differential access goals, such as the developers' ability to build and scale derivative products that address specific defender needs.
- **Drive collaborative product research and development:** Collaborate with cybersecurity companies, researchers, and governments to support the development and maturity of cyber-specific derivative systems and solutions. This could include providing earlier or exclusive access to new models.
- **Assess downstream capability changes:** Monitor how downstream developers are leveraging the model's capabilities to build cyber-specific applications and systems. This provides insight into emerging capabilities, allowing for adjustments to differential access implementation based on risk tolerance.
- **Provide technical assistance to downstream developers:** Help downstream developers leverage models by providing development guidance, educational resources, or other technical assistance.
- **Incentivize innovation:** Provide financial incentives or reduce costs for downstream development. This could include reducing licensing costs for any downstream developer or even setting up grant programs or competitive innovation challenges.

---

[49] The NASA Technology Readiness Levels (TRL) system lays out nine levels of technology readiness, which can be used as a framework to understand the development of AI-enabled derivative products (Manning, "Technology Readiness Levels.").

| Developer Strategies | Promote Access | Manage Access | Deny by Default |
|---|---|---|---|
| **Conduct Market Research** | Open industry engagement; public working groups; broad reports | Selective engagement with vetted partners; targeted research on critical gaps | Limited to internal research and discussions with trusted industry and government partners |
| **Prioritize Developer Access** | Open beta programs for all developers; prioritize inference capacity for those demonstrating innovation or security-related applications; minimal eligibility and screening requirements for model preview | Early access and increased inference limits based on developers' ability to build and scale derivative security products. | Highly selective early access only to developers serving highest-priority defense sectors; strict capability limitations; extensive security and scaling verification required |
| **Drive Collaborative Product Research and Development** | Open research collaboratives; broad early model access; shared testbeds | Selective research collaboratives; incubate specialized teams; co-develop critical infrastructure tools | Highly restricted partnerships; limited co-development. Consider only creating derivative security products in-house |
| **Assess Downstream Capability** | Minimal monitoring of derivative products; open innovation focus; | Regular assessment of downstream applications; adjust | Strict tracking of all derivative products or research; |

| **Changes** | minimal barriers | access based on capability development; Leverage licensing and use agreements to track developments | comprehensive review process; access revocation options |
|---|---|---|---|
| **Provide Downstream Developers Training and Guidance** | Public documentation; open resources; broad technical support | Tailored support for vetted partners; graduated access to advanced guidance; provide no-cost training to select downstream developers | Highly restricted documentation; intensive vetting before assistance; provide additional security training and guidance |
| **Incentivize Innovation** | Open grants; public competitions; reduced pricing and priority access for downstream security researchers, developers, and service providers | Targeted incentives for high-priority capabilities, significant cost reductions or grants for downstream security researchers, developers, and service providers | Limited incentives only for highest-priority applications with trusted development partners |

## 4.4 | Adoption, Scaling Up, and Iteration

Derivative product developers may need to adapt their initial prototype significantly to deploy it more widely in an enterprise setting, particularly in high-stakes settings like critical national infrastructure. As the prototype is applied in a real-world context, users will require that it be reliable, trustworthy, compliant with regulatory standards, and tightly integrated with their existing workflows. At the same time, when use of the product is scaled up, this can reveal underlying flaws. The developer may then need to further iterate on the product—for example, to address edge cases and missed detections, to improve capacity and performance, and to harden it against adversarial attacks.

### Developer Strategies

Foundation model developers can employ various strategies to support the adoption, scaling, and iteration of AI cybersecurity solutions by defenders across different sectors. These efforts will be dependent on both the differential access approach and the defenders. Key actions include:

- **Leverage a cyber capabilities-as-a-service model:** Provide cyber capabilities directly or through third parties, using a capability-as-service model. This reduces adoption and security burdens for downstream customers, enabling less mature or secure defenders to benefit from advanced capabilities while maintaining stronger control.

ASYMMETRY BY DESIGN | 43

- **Prioritize inference resources:** Allocate and manage computational resources strategically, providing preferential access and higher rate limits to defenders and service providers that contribute directly to differential access objectives.
- **Provide technical assistance to defenders:** Help defenders and scaling developers leverage models by providing comprehensive deployment guidance and educational resources. This supports proper implementation of controlled access systems and helps ensure defensive tools are used safely and effectively.
- **Incentivize defender adoption:** Provide cost reductions for licensing or inference costs to promote wider development and broader adoption. This particularly benefits critical but resource-constrained defenders such as CNI entities that may lack financial resources for advanced security solutions.

| Developer Strategies | Promote Access | Manage Access | Deny by Default |
|---|---|---|---|
| **Leverage a Cyber Capabilities-as-a-Service model** | Public APIs; flexible service offerings; minimal provider vetting; broad availability | Tiered service modes; provider vetting based on security practices; selecting service providers that scale up services and can implement access controls, monitoring KYC, etc. | Select service providers based on strategic priorities; consider in-house service development or partner with national governments to provide secure services to select defenders |
| **Prioritize Inference Resources** | Allocate compute resources and elevated rate limits specifically for selected defender groups or service providers | Strategically allocate compute for verified selected defenders or supporting service providers | Reserve rate limits and compute access for the most strategically important defenders or service providers, providing extensive vetting and ongoing assessment |
| **Provide Technical Assistance to Defenders** | Open documentation; public resources; broad support for all users | Tailored assistance based on defender verification; prioritized guidance for critical sectors; specialized support channels | Assistance to highest-trust defenders only; intensive security reviews; tightly controlled knowledge sharing |
| **Incentivize Defender Adoption** | Universal pricing discounts; open access programs; minimal qualification | Targeted incentives for verified defenders and service providers; further benefits based | Case-by-case support for essential defense entities; comprehensive evaluation process; |

|  | requirements | on ability to scale | strict usage requirements |

# 5 | Technical Infrastructure

Strategic implementation relies on technical infrastructure that enables fine-grained access controls and monitoring. **These controls determine *who* can access *what* capabilities under *which* conditions.** Some infrastructure already exists (e.g., rate-limiting less trusted actors), while other capabilities require further R&D. For example, methods to detect requests for dual-use capabilities at inference time could selectively allow only authorized actors to access certain capabilities.[50]

Technical infrastructure should not just be seen as an implementation detail—instead, it should be seen as a way to expand the range of possible differential access approaches (particularly in relation to Section 4.2, "Foundation Model Deployment"). These measures can be broken down into three categories:

- **Identity and authentication**: Mechanisms that verify user identity and manage their access credentials.
- **Model capability controls**: Mechanisms that manage how users interact with, deploy, or modify AI models and model outputs.
- **Monitoring and observability**: Mechanisms to track, log, and analyze user interactions with the AI system(s).

This section does not aim to provide a comprehensive taxonomy of tools and safeguards; instead, it aims primarily to lay out some possible options for better technical governance of differential access schemes, including directions for further research.

## 5.1 | Identity and Authentication

**Identity and authentication mechanisms** are used to ensure that access is only given to qualified and relevant parties (e.g., major cybersecurity providers that intend to leverage AIxCyber capabilities for defensive purposes).

At a high level, identity authentication and access management might involve several steps:

1. **Initial application:** A defender applies to access the developer's AI model or model outputs, based on basic or additional criteria outlined in Section 3.2. Higher access levels could require additional information, such as details on the intended use case for a deployed system and information on specific personnel who would be given access.

---

[50] This could build on existing proposals for structured access and safe harbor for evaluation and red teaming purposes. This proposal is also discussed in Reuel et al., "Open Problems in Technical AI Governance." p. 39.

2. **Authorization:** The developer, or a third party managing the process, then grants (or denies) access to a model or specific model outputs after conducting due diligence checks based on know your customer (KYC) processes. This could involve verification of an organization's legal status, assessment of their cybersecurity practices and internal controls, and background checks on specific personnel.
    - Access could require regular revalidation or be granted only until a specific use case is implemented (e.g., identifying and patching zero days for a specific software component).
    - For some arrangements, a third party could be employed to run the access management process (e.g., an independent body accrediting individuals as qualified cyber experts), rather than developers. This third party might be better suited to run this scheme due to their expertise or potential conflicts of interest (e.g., if the defender is a competitor).
3. **Authentication:** Access to the model or model output would require regular authentication through secure methods, including login credentials reinforced by multi-factor authentication (MFA) via hardware security keys. For higher-risk models, authentication becomes more restrictive—access may be limited to secured terminals within controlled facilities.

## 5.2 | Model Capability Controls

Model capability controls manage how users interact with, deploy, or modify AI models and model outputs. We split model capability controls for models with dual-use cyber capabilities into three main buckets (see table below).

### Approaches for Managing Model Capability Controls

In the context of differential access, developers should consider two technical levers: levers that alter the (1) general functionality of models, or (2) the cyber-specific capabilities of models.

Levers altering the **general functionality** of models can have a major impact on their overall usefulness, and can increase risks not just for dual-use cyber capabilities, but also for dual-use capabilities in other domains (e.g., biosecurity).[51] While more readily available, these levers can be cruder and cause unwanted side effects. Levers that affect **cyber-specific capabilities** selectively are less likely to impinge on overall model functionality and can provide narrower control over risks in the cyber domain, but some of these controls (e.g., developing input/output classifiers for dual-use capability access) may require further R&D.

---

[51] In addition to dual-use cyber capabilities, frontier AI systems can have a range of capabilities that can be misused or lead to accidents, Bengio, "International AI Safety Report 2025."

There are also some levers that are relevant for safety research and model interpretability but are unlikely to impact cyber functionalities directly, e.g., properties related to model inspection for structured access.[52] We consider these mostly not relevant to cyber capability controls.

## Categories of model capability controls

| Type of model capability control | Description | Examples |
|---|---|---|
| 1. General functionality | These are model properties that are likely to impact general capabilities if they are moderated. For example, instituting rate limits in LLMs would result in more limited capabilities across a broad set of domains. | Rate limits; limits on context windows; limits on inference time compute; allowing for model fine-tuning; restricting long-term memory |
| 2. Cyber-specific capabilities | These are model properties that could be managed to impact cyber capabilities specifically. For example, enabling tool integration or improving tool use specifically for cybersecurity tools, such as network and log analysis tools (Wireshark, Splunk). | Tool integration; agent-to-agent calls; command line access; access to specific cybersecurity capabilities (e.g., via input/output classifiers[53]) |
| 3. Model properties that are unlikely to directly affect cyber capabilities | These are model properties that, even if access is given, are unlikely to directly affect cyber capabilities and therefore likely do not need to be controlled in the context of the differential access framework this report focuses on. There could be other reasons to limit or control access to these properties though (e.g., to prevent trade secret leakage). | Ability to view output logits and probabilities for a given input; ability to inspect model parameters and activations[54] |

---

[52] Bucknall and Trager, "Structured Access for Third-Party Research on Frontier AI Models."
[53] Sharma et al., "Constitutional Classifiers."
[54] Many properties that are useful for safety evaluations and safety research (e.g., on model interpretability) are less relevant for enabling defensive applications. There is more detail on alternative access management schemes in Bucknall and Trager, "Structured Access for Third-Party Research on Frontier AI Models."

We suggest that developers consider two approaches to managing model capabilities:

1. **Defensive acceleration**: The developer proactively provides resources and AI capability management beyond baseline access to specific defenders, aiming to encourage developing and using defensive AI applications.
2. **Conditional capability extension:** Upon request by a defender, the developer grants additional access to capabilities and control of model properties beyond baseline access for specific, justified use cases.

A **defensive acceleration** approach can be used when there is a model where there are no significant restrictions for regular uses, but there are still potential defensive applications that could be incentivized among defenders. In this case, a developer can provide additional resources, such as increased compute allocation and higher rate limits for security-specific tasks. Dedicated technical support could also be provided for using the shared AI system in defensive workflows or for helping improve scaffolding to elicit defensive capabilities better.[55]

**Conditional capability extension** can be used when a model has specific properties restricted due to a significant risk of misuse. A defender may have identified a valuable defensive application that can be performed if some restrictions are eased; for example, by allowing integration with a new tool. The developer can agree to grant additional access after evaluating the validity of this use case, possibly requiring additional monitoring requirements for defenders to use these less restricted systems (see [Section 5.3](#) for more discussion).

However, conditional capability extension is a more attractive option in relation to cyber-specific capabilities than for general functionality of models. While the developer can constrain the general functionality of a model to limit the unnecessary proliferation of dangerous non-cyber capabilities, doing so will also limit the model's utility for legitimate defensive use cases.

## Cyber-Tool Provision

One alternative approach to restrict the risk of misuse from a general-purpose model could be **cyber-tool provision**. Instead of providing access to the base general-purpose model, the developer can provide defenders with a narrow version of the model whose usefulness is limited to a particular set of cybersecurity applications.[56]

---

[55] Offensive cybersecurity capabilities in current LLMs are likely to be under-elicited, with researchers managing to achieve significant performance improvements (95% vs. scores of 72% and below on intercode-CTF) via improving basic scaffolding components and prompting strategy, and including standard Linux tools and Python packages (Turtayev et al., "Hacking CTFs with Plain Agents.").

[56] For example, other researchers have used the term "tool AI" to refer to a type of AI system that is built to have a more narrow set of capabilities and goals, rather than to act autonomously in the world as an agent (Bengio et al., "Superintelligent Agents Pose Catastrophic Risks.").

For example, if a model presented misuse risks in multiple domains (e.g., biosecurity and cybersecurity), the developer could instead provide a narrow model focused solely on software engineering and cybersecurity. While a malicious actor gaining access to the model could use it to perform malicious cybersecurity activities (or some subset thereof), they would not be able to use it to inflict other potential harms.

Cyber-tool provision requires technical methods to limit the model's capabilities and behaviors. Some possible technical approaches could include:

1. **Model distillation:** Smaller models could be distilled to replicate behaviors of the larger model that are specifically helpful for a particular defensive cybersecurity function. This model could also be fine-tuned further on use-case-specific tasks.
2. **Circuit breakers and "unlearning":** internal representations related to non-intended uses of the model are disrupted so that the model cannot generate outputs unrelated to the primary use case.[57]
3. **Refusal training:** A technique where an AI model is specifically trained (usually through fine-tuning) to recognize and decline to engage with certain types of requests or to avoid generating certain categories of outputs.
4. **Input/output filtering:** A classifier could be used to screen inputs or outputs before an output is returned to the user. These would not alter the model's underlying capabilities.

Refusal training and input/output filters are vulnerable to jailbreaks, potentially making model distillation and unlearning more attractive for reducing proliferation risk. However, more research is needed to demonstrate if these methods can successfully limit a model's capabilities while leaving it sufficiently useful for narrow tasks.

## 5.3 | Monitoring and Observability

Monitoring and observability mechanisms help developers to track, record, and analyze how users are interacting with an AI system. This allows oversight of how the shared model or model output is being used, enabling detection of unauthorized use, and also provides data for improving how access control mechanisms work over time.

While implementing monitoring and observability mechanisms is a core part of any well-functioning differential access scheme, there is a balance to be struck in maintaining sufficient visibility while not unduly creating a barrier to adoption by legitimate defenders. Defenders might have reasonable desires to maintain confidentiality around their clients or their own organizational security, may want to protect intellectual property, or may have to comply with legal obligations to protect customer data.

---

[57] More research is needed to determine whether this class of safeguards is appropriate for this particular use case. For more on circuit breakers, see Zou et al., "Improving Alignment and Robustness with Circuit Breakers."

To access higher-risk models with higher capability levels, developers can require users to opt-in to enhanced monitoring. However, this creates a potential conflict: what is best for society (having skilled defenders use these powerful AI tools with appropriate oversight) might not align with what individual defenders or organizations are willing to accept in terms of monitoring.

Developers may need to carefully calibrate monitoring requirements to balance security oversight with practical usability. To enable risk management, developers should strongly consider monitoring:
- **Basic usage patterns**, such as volume and frequency of requests, types of API endpoints accessed, time patterns of system usage, and resource consumption
- **Authentication data**, such as login attempts, access key usage, and session patterns, can help identify unauthorized access attempts and does not require disclosure of potentially sensitive content
- **Model interaction information**, such as input/output patterns, reasoning tokens, and tool integration requests. These can reveal attempts to use the model in unauthorized ways, though input and output information could contain confidential data the defender may be concerned about sharing.

Defenders are likely to want to protect the following types of information from being shared and stored:
- **Security infrastructure details**, which defenders would want to keep private because their exposure could provide attackers with valuable intelligence about defensive capabilities and vulnerabilities
- **Confidential business information**, which could include customer data, financial data, and intellectual property
- **Personally identifiable information**, including information that could be used to identify specific persons, such as name, government ID, biometric records, etc.

There is a tension between security and privacy: some information that may be useful for assessing the unauthorized use of models or model outputs is also information that defenders would find too sensitive to share outside their organization. While some of this can be resolved bilaterally between developer and defender parties, it is possible that technical solutions would allow for better

ASYMMETRY BY DESIGN | 51

monitoring and privacy.[58] Various privacy-enhancing technologies (PETs) could be used to reduce the privacy-security tradeoffs involved in monitoring within a differential access scheme.

- PETs are various mechanisms that allow for information-sharing between parties while enforcing precise information flows. Technologies such as homomorphic encryption, differential privacy, and zero-knowledge proofs can enable input/output privacy, input/output verification, and flow governance.[59] For example, homomorphic encryption could allow defenders to encrypt their sensitive data (such as customer data) while still enabling monitoring and analysis on that data without it being decrypted, specifically by analyzing broader usage patterns.
- Concerns around storage of sensitive data can be managed through the use of secure enclaves, which are a protected area within a computer processor that operates independently from the main system, creating a secure environment for processing sensitive data.[60] These enclaves use hardware-based isolation and encryption to ensure that even if the main system is compromised, the data and operations within the enclave remain protected.

---

[58] It is worth noting though that there is a commercial incentive between competing AI developers to undercut on security, e.g., by giving access with less monitoring or data retention, because users may prefer developers that prioritize their privacy.
[59] For more on structured transparency as a framework to deal with trade-offs involved with information-sharing, see Trask et al., "Beyond Privacy Trade-Offs with Structured Transparency."
[60] For more on using secure enclaves in an AI evaluation context, see Trask et al., "Secure Enclaves for AI Evaluation."

# 6 | Promising Schemes

The following examples bring together all framework elements in real-world schemes tailored to specific challenges. Effective differential access schemes address particular problems rather than applying generic solutions. These examples demonstrate how different organizations implement differential access for distinct cybersecurity challenges.

- This problem can be a **threat scenario for a specific user group** that is created by novel capabilities or misuse thereof (e.g., threats to CNI owners and operators).
- It can be a more **general problem relating to a set of capabilities**, such as how to prevent an "offensive overhang" from emerging at high capability levels if the government elects to tightly restrict access to advanced systems.

As illustrated in the following examples, there are a large number of possible schemes based on what capabilities, defenders, technical mechanisms, and organizational strategies one uses. By focusing on a specific problem, it becomes easier to narrow down and set these parameters and design a scheme that suits a particular organization.

**The following section is not an exhaustive list of all possible schemes, nor does it decisively recommend a top candidate.** Instead, it presents illustrative models and their costs and benefits. There are numerous threat scenarios and capability developments that are not covered. For instance, differential access may also be helpful in addressing threats to the software supply chain of AI labs or the defense-industrial base, which we do not address.

This section outlines four sample differential access schemes, showcasing a range of designs tailored to different capability levels and relying on different technical and policy levers:

| Name of Scheme | Type of Access | Description |
| --- | --- | --- |
| Scheme A: Accelerator for CNI Cybersecurity Innovators | Promote Access | Early access and policy incentives (e.g., financial) for large number of startups and established product vendors to incentivize AI-based solutions for CNI security (e.g., better threat detection, identifying misconfigurations) |
| Scheme B: Dual-Use Authorization for Security Researchers | Manage Access | Expanded access to dual-use capabilities (e.g., exploit generation) for large number of white-hat hackers to improve the security of the open-source community, using technical infrastructure to manage and monitor their access |

| Scheme C: Rapid Response Force of Keystone Defenders | Manage Access | Foundation model developers convene a smaller, trusted group of keystone defenders to reduce vulnerabilities in software and improve threat detection, focusing on rapid experimentation and prototyping of new AI capabilities |
|---|---|---|
| Scheme D: High-Capability Adversarial Testing as a Service | Deny by Default | Foundation model owner/operator (e.g., government) tightly controls access to highly capable system (access for very few users) but conducts high-end penetration testing and red teaming for other actors, providing automated "nation-state attack emulation as a service" |

These schemes are not mutually exclusive; there may be foundation models of varying capabilities available in the world. Medium-capability systems could be made more widely available to a large number of users via Scheme B ("dual-use authorization for security researchers"), while other higher-capability systems could be limited to a much smaller select group via Schemes C or D.[61]

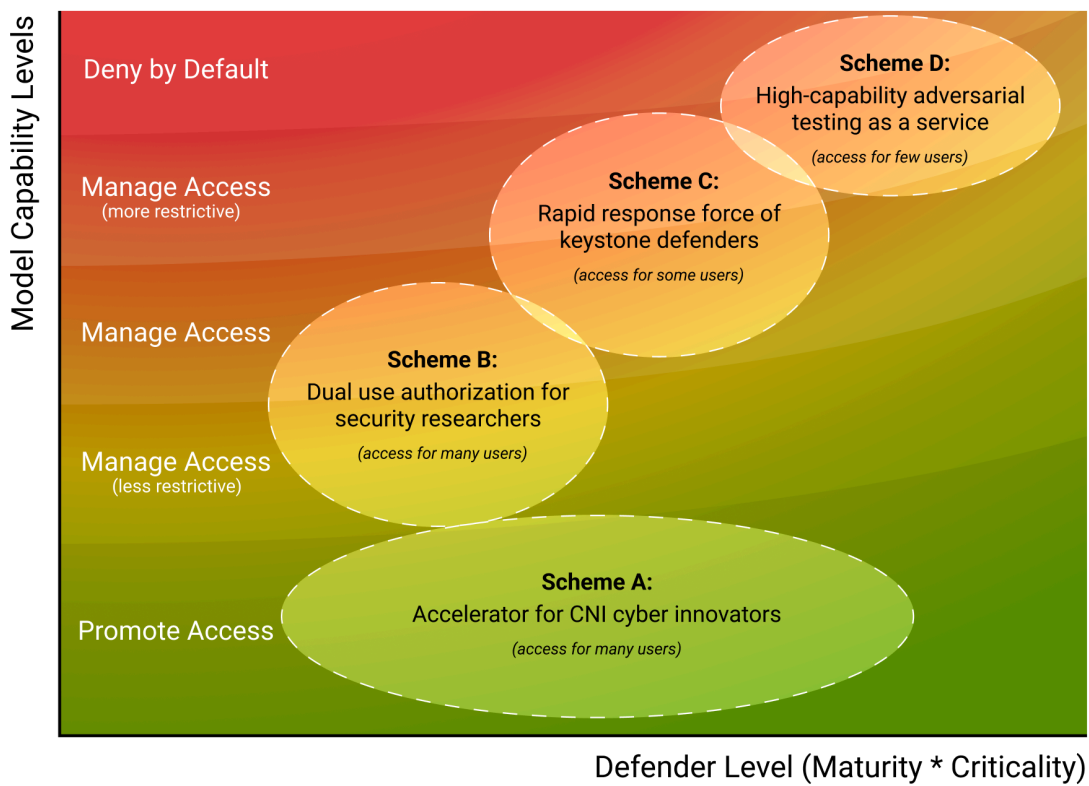

Note: This diagram is illustrative. Schemes are depicted as ellipses for ease of reading, but need not conform to this (and would likely be substantially more overlapping than depicted in this chart).

---

[61] We mention the number of users here as a heuristic for the degree of access allowed, but this need not map precisely to the restrictiveness of the scheme. Developers may also want to grant access based on other criteria, e.g., number of organizations, rather than number of users.

## 6.1 | Scheme A: Accelerator for CNI Cybersecurity Innovators

**The problem:** Due to challenges encountered across the product development lifecycle, cybersecurity product vendors in the energy sector (and other actors) require substantial lead time to prototype and scale AIxCyber tools that respond to new capabilities.

**The solution:** This aligns with a Promote Access approach, where (1) frontier model developers provide CNI cybersecurity product vendors and startups with early access and expanded access to LLMs/reasoning models; (2) policymakers drive CNI adoption by providing support over the product development lifecycle such as convening, testbeds, regulatory sandboxes, and improving AI literacy among CNI owners and operators.

| What capabilities? (Section 2) | Who gets access? (Section 3) | What strategic considerations? (Section 4) | What technical infrastructure? (Section 5) |
|---|---|---|---|
| Varied—e.g., threat detection; identifying misconfigurations; asset inventory | Energy sector cyber defenders—e.g., established product vendors, startups<br><br>*(access for many users)* | Broad support for derivative product development from foundation model developers and policymakers | Minimal—some KYC; possibly expanded access to AI models |

*Interested in reading more on elements of this scheme? See: Section 1.3 "Promote Access"; Section 3.4 "Low-Maturity Critical Actors"; Sections 4.3, 4.4 on derivative product development*

**More detail:** For many CNI actors (e.g., energy sector owners and operators), access to AI capabilities alone is not enough. These actors often have limited AI talent and cybersecurity resources, and their systems are safety-critical and subject to heavy regulatory oversight.

Effective support for these actors requires targeting cybersecurity product vendors that support them, as well as encouraging adoption among larger CNI owners/operators and MSSPs that support smaller CNI owners/operators. Applications are varied and could include improved threat detection, asset inventory, or misconfiguration identification for CNI; even identifying the most promising gaps and applications is a challenge for industry and policymakers to solve.

In terms of differential access levers, this is largely a policy problem rather than a technical problem; again, the primary bottlenecks are around the product development and adoption lifecycle, rather than novel technical methods to control/promote access. Potential levers include:

1. **Developers can provide expanded AI access** to CNI cyber product vendors and startups, which could include:
    a. *Early access*, in the form of granting some actors additional time to experiment with applications of LLMs/reasoning models in CNI cybersecurity defense;
    b. *Expanded access*, e.g., higher rate limits or context windows, allowing for fine-tuning on cybersecurity-relevant datasets, etc.
2. **Policymakers can provide support** to established CNI cyber product vendors and younger startups, which could include interventions in phases such as:
    a. *Identifying use cases*—for example, by facilitating working groups or commissioning reports about applying LLMs/reasoning models in CNI cybersecurity defense;
    b. *Prototyping*—for example, by creating public resources like OT/ICS testbeds where startups can test their products either in simulated environments (such as digital twins of systems) or on physical infrastructure;
    c. *Scaling up*—for example, by creating regulatory sandboxes for companies to test out pilot programs or improving literacy among CNI owners and operators (and other actors in the ecosystem).

## 6.2 | Scheme B: Dual-Use Authorization for Security Researchers

**The problem:** Cybersecurity capabilities are dual-use, and restricting them broadly could penalize a large community of white-hat hackers that otherwise could improve the overall security, including open-source software security.

**The solution:** This aligns with a Manage Access approach, where developers use technical methods (e.g., input/output classifiers) to filter out requests for dual-use capabilities (e.g., developing proof-of-concept exploits) for most users, and only allow authorized users to request these capabilities.

| What capabilities? (Section 2) | Who gets access? (Section 3) | What strategic considerations? (Section 4) | What technical infrastructure? (Section 5) |
|---|---|---|---|
| Exploit development and vulnerability discovery | Authorized white-hat hackers (e.g., via a bug bounty program) and academia  (access for many users) | Tiered access for verified defenders, primarily around early access and foundation model deployment | Filtering of dual-use capability requests; KYC measures and authentication; increased logging and monitoring |

*Interested in reading more on elements of this scheme? See: Section 1.4 "Manage Access"; Section 3.5 "Force Multipliers"; Sections 5.1-5.3 on "Technical Infrastructure"*

**More detail:** AI developers may want to restrict access to some dual-use AI capabilities, such as the ability to develop proof-of-concept (PoC) exploits. While a PoC exploit can help defenders prioritize fixing a specific vulnerability among numerous potential issues, malicious actors could also incorporate these exploits into their malware, causing real-world harm.

Independent security researchers (e.g., "white-hat hackers") and academia play an important role in securing the digital ecosystem, especially when it comes to securing open-source software. Vulnerabilities in open-source software can compromise the software supply chain of top AI labs and others.[62] Allowing this community, potentially comprising tens of thousands of skilled individuals, to access cutting-edge cybersecurity tools could meaningfully improve overall security.[63]

One approach could be for AI developers to implement a filtering system for dual-use cyber capabilities like PoC exploit generation. Under this approach, **the majority of users would have access to a base model** that automatically blocks requests for dual-use cyber capabilities. In contrast, **a subset of trusted researchers would have access to more advanced capabilities** that would require user authentication. This could involve requiring these researchers to undergo a KYC process during signup and to authenticate themselves whenever they intend to use the advanced features.

Implementing this would require developing (1) technical safeguards to accurately classify requests and resist potential jailbreaks and (2) a system to identify and authenticate trusted researchers and monitor their use of dual-use capabilities. Given the wide range of potential users, AI developers might consider requiring trusted researchers to agree to more granular activity logging and measures such as MFA. This would enable developers to identify and prevent potential misuse of dual-use capabilities, such as those resulting from credential theft and abuse.

Policy interventions may also be needed to establish "rules of the road" for capabilities to which developers can reasonably provide expanded access. Developers might hesitate to broadly grant users access to potentially harmful cyber capabilities without explicit policymaker approval, due to liability concerns. Governments could consider granting safe harbor to developers for a dual-use authorization scheme that covers a prescribed range of capabilities and is conditional on developers implementing appropriate safeguards, and to independent researchers for conducting security research using developers' AI tools.

---

[62] Druttman, "Breaking Down Nation State Attacks on Supply Chains."
[63] e.g., HackerOne has over 1 million registered users, though probably most of these are not active or are hobbyists and less likely to be impactful, HackerOne, "HackerOne Reveals Industry and Company Growth as Enterprises Secure Rapid Digital Transformations."

## 6.3 | Scheme C: Rapid Response Force of Keystone Defenders

**The problem:** As AI capabilities advance, attackers will be able to move more quickly through their OODA (Observe, Orient, Decide, Attack) decision-making cycles. This acceleration will increasingly render ineffective the traditional software security model—where developers release software, identify vulnerabilities, create patches, and rely on users to apply them—particularly when products have millions or billions of users.

**The solution:** This aligns with a Manage Access approach, where frontier AI developers can convene a rapid response force of Keystone Defenders—mature organizations that play an outsized role in the digital or cybersecurity ecosystem—to adopt a secure-by-design strategy. These organizations would proactively use AI systems to detect compromises within their own systems and to identify software vulnerabilities before release, thereby shortening or bypassing traditional patch management cycles.

| What capabilities? (Section 2) | Who gets access? (Section 3) | What strategic considerations? (Section 4) | What technical infrastructure? (Section 5) |
|---|---|---|---|
| Varied—vulnerability discovery; faster patching; threat detection/response | Keystone Defenders *(access for some users)* | Convening small, trusted coalitions for experimentation and rapid prototyping of AI-enabled tools | KYC; authentication; possibly throttling of unneeded skills (e.g., bio) for highly capable models |

*Interested in reading more on elements of this scheme? See: Section 1.4 "Manage Access"; Section 3.3 "Keystone Defenders"; Sections 4 and 5 generally*

**More detail:** Even if the use of AI in cybersecurity is ultimately defense-favored, the question in the near-term is one of speed. Once vulnerable software is released, the lifecycle for patching it can be protracted. At best, it can take weeks or months if developers can fix the vulnerability quickly[64] and users act fast to apply the patch.[65] At worst, it can take years, with some vulnerabilities remaining undiscovered and unpatched while sophisticated actors like nation-states continually exploit them.[66]

---

[64] Morrow, "Patch Now: Vulnerabilities Exploited in 5 Minutes?"; Mitesh, "Patch Tuesday Turns 20."

Advances in AI capabilities could decrease attackers' time-to-exploit while also expanding the number of vulnerabilities that they can find. This makes a secure-by-design approach critical for Keystone Defenders in the software supply chain—e.g., actors like Apple, whose devices are used by billions of individuals globally; or security providers like Crowdstrike, who are trusted to provide services to tens of thousands of enterprise users. These actors will also need to guard against their own systems being compromised.

Foundation model developers could build small, trusted coalitions to focus on the use of AI to accelerate software testing and patching and to improve threat detection and response, with an emphasis on experimentation and rapid prototyping of solutions. These coalitions could involve governments, as well as industry organizations and civil society members like the Open Worldwide Application Security Project or the Linux Foundation. To improve the agility of Keystone Defenders, governments might also consider providing funding to support these collaborative efforts, or removing barriers that impede timely collaboration among leading organizations—for instance, by offering safe harbor protections against potential antitrust actions or other legal risks, though such protections will have to be scoped carefully so as not to unduly advantage incumbents.

Technical infrastructure could also contribute to the success of this scheme. Rapid response force members may need ways to avoid revealing proprietary information to each other. They may also need to submit to authentication and monitoring measures as outlined in Scheme B. Lastly, the foundation models they use may be highly capable general-purpose AI systems, with dual-use capabilities not just in cybersecurity, but in other domains (e.g., biosecurity, weapons development and acquisition). If so, the rapid response force may prefer to limit or remove non-cybersecurity capabilities (a "cyber-tool provision" approach as outlined in Section 4.2) to minimize risks from bad actors, such as insider threats acquiring dual-use capabilities in these other domains.

---

[65] The time taken to apply patches after their release can range from a few days to several months, depending on the specific circumstances and organizational practices. In some cases, and with some organizations, automated patch management is an option, deploying patches as soon as they are released by developers (Paliwal, "How Automated Patch Management Improves IT Security and Efficiency."). It is also worth noting that other faster-to-deploy options may exist; e.g., configuration changes, network controls, and other compensating controls could provide the buffer that would allow a patch to be deployed at a later date.

[66] Famously, EternalBlue was used by the NSA to exploit systems since at least 2012, before being publicly disclosed (and subsequently fixed) in 2017. Source: Nakashima and Timberg, "NSA Officials Worried About the Day Its Potent Hacking Tool Would Get Loose. Then It Did."

## 6.4 | Scheme D: High-Capability Adversarial Testing as a Service

**The problem:** At extremely high capability levels, developers may be unwilling to share access to the foundation model except perhaps with a small trusted group of cleared individuals. However, this risks creating an "offensive overhang" where defenders remain vulnerable to other sophisticated adversaries that may develop or acquire their own advanced cyber capabilities.

**The solution:** This aligns with a Deny by Default approach, where owners/operators of foundation models could offer services to Keystone Defenders without granting direct access to an advanced capability. For example, owners/operators could share information about novel vulnerabilities their models have uncovered or offer automated penetration testing that emulates nation-state attacks to help Keystone Defenders strengthen their own systems.

| What capabilities? (Section 2) | Who gets access? (Section 3) | What strategic considerations? (Section 4) | What technical infrastructure? (Section 5) |
|---|---|---|---|
| Very advanced capabilities—e.g., full automation of cyber operations and novel attack techniques | Direct access very limited; defenders can receive indirect access as a service *(direct access for few users)* | Primarily "Deny by Default" approach | Strict requirements around identity and authentication and monitoring |

*Interested in reading more on elements of this scheme? See: Section 1.5, "Deny by Default"*

**More detail:** In certain scenarios, a foundation model owner/operator may wish to take a Deny by Default approach to their general-purpose AI system, broadly restricting access because its high capability levels pose a danger to the public, and existing safeguards do not yet provide sufficient protection against potential misuse. While hypothetical, such a scenario would pose a significant challenge for applying the system defensively.

One workaround may be for the owner/operator to provide these capabilities as a service without granting access to the foundation model. Some examples could include:

- **Building on existing vulnerability disclosure policies** (e.g., the Vulnerabilities Equities Process for the U.S. government, or responsible disclosure practices more generally for companies) to notify affected parties if a vulnerability is identified in a system they have developed. Disclosure of information would have to be conducted carefully, and could be limited if the vulnerability is difficult to fix (e.g., for side-channel attacks) or there are other national security interests at play.
- **Offering adversary emulation campaigns**, such as simulating an attack by a top-tier nation-state to identify gaps in an organization's defenses. Adversary emulation can take weeks or months, and there is a limited number of highly skilled offensive cybersecurity professionals. Being able to deliver nation-state-grade adversary emulations at scale could help defenders in the private sector get ahead of attackers seeking to develop similar systems. The owner/operator would have to carefully balance transparency against operational security, and may only be able to release redacted "lessons learned" rather than details of the attack techniques used.

However, establishing such a service may require considerable effort and resources. Realistic adversary emulation exercises might involve granting the foundation model owner/operator access to sensitive systems of Keystone Defenders or critical infrastructure, which would require managing risks to privacy, intellectual property, safety, and security. If the government is involved (e.g., intelligence community), users of this service may also be concerned about weaponization of vulnerabilities discovered in classified systems, or regulatory impacts such as liability for software defects. And if the foundation model developer is operating this service on their own, then they may not be incentivized to shoulder the upfront costs of setting up vulnerability detection or simulation services unless their pricing model reflects that, potentially making the cost of such a service prohibitive to actors that need the service the most.

# 7 | Future Work

As AI model capabilities advance, discussions regarding how to prevent their misuse while enabling beneficial applications will continue to evolve. This report represents an initial framework for assessing and implementing a differential access approach. Significant additional research is necessary in multiple areas—particularly in threat modeling. There remains a pressing need for deeper, more structured threat modeling work; without rigorous, domain-specific models, it is difficult to justify or calibrate levels of differential access.

The recommendations in this report serve as starting points for community exploration rather than as specific models for differential access. We encourage others in both the cybersecurity and AI communities to continue to explore differential access approaches and implementation schemes. This includes not only the developers themselves, but academia, government, and independent organizations. As risks and capabilities evolve, differential access mechanisms aided by this and future research may become necessary not only for preventing misuse, but for ensuring that advanced cyber capabilities support legitimate actors.

## 7.1 | Recommendations and Further Research

### General Recommendations

AI companies play a vital role in research and development. To manage risks from model deployments, most leading AI companies now have "frontier AI safety frameworks" or "responsible scaling policies."[67] AI companies can consider incorporating differential access schemes for when systems need additional safeguards, e.g., via:

- **Capability and risk thresholds:** First, companies could address and prepare for "trigger points" in their safety frameworks that would require incorporating differential access approaches. AI companies could stipulate the need to implement certain differential access schemes at a given AIxCyber capability threshold, such as Google DeepMind's Cyber Uplift Level 1 threshold, which involves a system that could help "well-resourced threat actors carry out severe cyber attacks" (e.g., on critical infrastructure).[68]
- **Capability and Risk Assessments:** Companies could prepare for both unexpected occurrences such as newly discovered offensive capabilities, findings from security audits, or shifting user bases or priorities, and also develop specific, regularly-occurring checkpoints when access policies are reassessed.

---

[67] Frontier Model Forum, "Issue Brief."
[68] Google DeepMind, "Frontier Safety Framework v2.0."

- **Risk Governance:** Company safety frameworks could also develop mechanisms for escalation, should triggers or checkpoints generate concerns regarding current levels of access. For example, a cross-functional access review body could exist to assess these changes and determine follow-up actions. Members of this team could include legal representatives, model alignment developers, and relevant cybersecurity experts. In severe, high-risk or urgent cases, safety frameworks could have an additional escalation option in which immediate interim action such as temporarily disabling access could occur, giving the committee time to review the circumstances.

Additional research could also involve conducting comparative analyses of past security breaches that could have been mitigated or exacerbated by AI-enhanced defensive or offensive cyber tools.

## Threat Modeling, Assessing Defenders, and Strategic Considerations

*Threat Modeling*

Threat models can help decision-makers identify whether a capability introduces plausible, realistic threats—and under what conditions those threats make a serious impact. Robust threat modeling can inform and justify implementing differential access. Further, this can help justify the implementation of differential access schemes, as these schemes can be costly for the frontier AI developer, and can potentially disadvantage stakeholders in the cyber community that do not get access.

AIxCyber threat models should provide information including:
- Assessments on the real-world feasibility and cost of misuse;
- Estimations of threat actor motives, incentives, and capabilities; and
- Development of concrete, plausible scenarios for misuse.

Governments, companies, and researchers should collaborate to analyze high-impact cyber threat scenarios that might pose the greatest risk over the next 10 years. While this report identifies some preliminary threat models, including AI-enabled worms or attacks on critical national infrastructure, detailed threat modeling is not our main focus.

Other threat models worth considering may include (non-exhaustively):
- Protecting military capabilities, such as the defense industrial base, in the event of a great power conflict
- Protecting the AI hardware/software supply chain, including AI labs, data centers, and related infrastructure

*Assessing Defenders*

Future research could also involve providing greater detail around potential users of differential access. The Manage Access approach rests on identifying tiers of users in detail (see Section 3), while the Promote Access approach requires less detail, but still involves identifying certain actors for privileged access. Concrete implementation of these approaches would require principled guidelines that allow frontier AI developers to differentiate between user tiers, and concrete assignment of organizations/individuals to each tier. Further research should focus on building out such guidelines and test-running them with actors who might receive access.

This work could also focus on:
- developing specific KYC criteria for these user groups or capability levels;
- mapping the software supply chain in greater detail to understand which (types of) actors are most important;
- validating barriers to access for Keystone Defenders, as the market may naturally lead to socially optimal outcomes;
- identifying priority applications of AI for actors protecting CNI entities.

Scenarios that involve a Deny by Default approach might require exploring how to provide controlled disclosure of important information for chosen actors. Further research here could involve exploring what scenarios might require governments or companies to restrict access significantly, and laying out a spectrum of possible opportunities to manage the disclosure of sensitive information. For example, if a new class of exploits were discovered, options could include sharing, e.g., the proof-of-concept, full research methodology, workable code, or discreetly notifying the owners of a vulnerable system that their system could be affected.

*Other Strategic Considerations*

Differential access schemes would require different amounts of administration and oversight. Further research could involve examining what resources would be necessary to sustainably implement a scheme, including budgetary and logistical factors, staff needs and oversight, and additional staff specialization and training. Such research could also examine the potential economic costs/benefits of implementing differential access.

Researchers could also assess how attackers will respond to and adapt to differential access schemes, using game theory and other methods. Understanding the net effectiveness of differential access schemes will require knowledge of whether potential malicious actors will choose to invest in AIxCyber capabilities to acquire these capabilities by hacking the chosen defenders, or to respond by other means.

## Evaluations and Technical Infrastructure

*Capability Evaluations*

Differential access could potentially be used to gather information about AI cyber capabilities, by giving relevant groups early access to the model and also incentivizing them to experiment with the model and its capabilities. This could address current issues with the under-elicitation of cyber capabilities, such as limited time/resources for in-house evaluators and limited access for third-party evaluators like METR. Further research could explore how to use differential access not just for improving cyber defenses, but also for improving capability evaluations.

AI Safety Institutes (AISIs), researchers, and companies should also consider developing capability evaluation suites for defensive cyber capabilities, as current evaluations focus primarily on offensive cyber capabilities. This work could also:

- validate and/or build on the proposed taxonomy provided in Section 2 of this report (which in turn builds off Pattern Labs's taxonomy of offensive cyber capabilities);
- investigate foundational capabilities that might underlie both offensive and defensive capabilities (e.g., via factor analysis to identify underlying clusters of capabilities);
- map AI capabilities to existing cybersecurity frameworks, such as the MITRE ATT&CK, MITRE D3FEND, or OWASP threat categories.

*Technical Infrastructure*

Future R&D around new technical infrastructure could enable more flexible differential access schemes. Some examples could include:

- Research to restrict or throttle dual-use capabilities so that developers can control special access to cyber capabilities for authorized users. For example, classifiers for dual-use cyber capability requests could allow "trusted researchers" to access capabilities only if authorized to do so, or develop more fine-grained tiers for user requests based on the likelihood/impact of a dual-use request being used maliciously.
- "Cyber-tool provision" research to restrict general-purpose models to only specific (defensive) cyber functions to reduce the overall risk of providing access to a particular AI system (for highly capable systems where the risk may stem not just from cyber, but e.g., biosecurity concerns, weapons design). Such techniques could include model distillation, "unlearning" or circuit breakers, or refusal training.
- Ablation studies on cyber agents performing various tasks, both for (1) model properties affecting general capabilities (e.g., rate limits, context window limits, etc.) and (2) model properties that affect cyber-specific capabilities (e.g., tool integration, command line access). This could be used to determine what capabilities provide the most powerful levers for controlling (or boosting) cyber agents.

While the above controls focus primarily on proprietary models, open-weights models such as DeepSeek have, in some cases, been highly competitive with proprietary frontier models. Future differential access schemes may require technical infrastructure not just from model developers, but also from inference providers.

## Government Policy

Governments should consider how to inform, incentivize, and support differential access strategies—in particular, by implementing policies that support developers and defenders at different stages of the model and derivative product lifecycle. Government intervention can be effective at both early and late stages of the product lifecycle. Early intervention could include investments in basic research and proof-of-concept projects similar to the Defense Advanced Research Projects Agency (DARPA)'s Artificial Intelligence Cyber Challenge (AIxCC). At later stages, when mature products exist but face adoption hurdles, government incentives targeting specific use cases can accelerate implementation by reducing financial, regulatory, and expertise barriers.

- **Investing in research and development**, e.g., through supporting early research, funding pilot projects, or creating innovation challenges focused on developing AIxCyber capabilities for high-priority security applications or specific defenders
- **Removing regulatory barriers**, e.g., through regulatory sandboxes, particularly for highly regulated industries (e.g., critical infrastructure)
- **Reducing costs for using these systems**, e.g., via subsidies or financial incentives, such as providing tax incentives for CNI owners/operators to adopt tools that automate misconfiguration detection in their IT environments

*Differential Access and Market Competitiveness*

Differential access approaches that favor specific defenders may create competition and antitrust issues. These issues extend well beyond the scope and expertise of the authors, but we believe this is an area that requires further study. As with many areas, government policymakers will need to balance national security concerns with market competitiveness.

## 7.2 | Limitations

Differential access in the AIxCyber space is not without its challenges and limitations. The cybersecurity community has historically valued openness as a key element of innovation and may not support differential access; differential access schemes need to be motivated by concrete, realistic threat models; differential access may not be useful in certain scenarios; and, finally, in some cases, tolerating certain offensive cyber attacks may be a strategic advantage for states.

First, openness has historically been an important norm in the cybersecurity community, both for pragmatic reasons and due to cultural overlap with the open-source community.[69] Broad access can permit independent researchers—who may not have formal affiliation with an institution—to still make significant research contributions. It may also permit better peer review and community scrutiny, allowing vulnerabilities to be caught and fixed earlier if more parties have access to and can inspect code. Limiting access to AIxCyber tools to large organizations or governments may also create a power imbalance by excluding smaller stakeholders. Some companies and security researchers may hence be skeptical of a system that privileges access for some actors over others. To ensure buy-in from the cybersecurity community, companies implementing differential access schemes should strive to be transparent about the heuristics they use to grant access and ensure that limitations on broad-based access are justified effectively.

Second, differential access schemes require a compelling motivation to implement—i.e., they should address a tangible and significant risk. Currently, there is still significant uncertainty around what the biggest AI-enabled cyber threats are, and much more work needs to be done to explore such threat models. Identifying specific AI-enabled threats is beyond the scope of this report, but significantly more threat modeling work will be necessary to identify which differential access schemes are most appropriate.

Differential access also may not be useful in all cases. AI problems do not always necessitate AI solutions, and in some cases, the best solution will be a tried-and-tested method. For example, promoting security by design—i.e., ensuring that developers avoid introducing vulnerabilities in the first place—remains one of the most effective ways to pre-empt security issues. Other high-leverage security practices, such as multi-factor authentication and zero-trust architecture, will remain invaluable regardless of AI progress. And as has been true for many years already, effectively securing critical infrastructure will require a concerted effort to invest more in the security of under-resourced defenders (e.g., utilities).

Lastly, from the view of some international relations theorists, it may sometimes be strategic for states to tolerate various kinds of cyberattacks, for reasons that might include deterrence,[70]

---

[69] The cultural overlap between the cybersecurity and open-source communities has been a key driver supporting openness for decades. For example, in 2008, Ross Anderson wrote: "In the long run, openness improves security. This is a lesson the open-source movement learned early, and one that the security community has taken increasingly to heart." (Anderson, *Security Engineering: A Guide to Building Dependable Distributed Systems*.) Ross Anderson, Security Engineering, 2nd ed. (Wiley, 2008), Chapter 20, Section 20.4.2. Further back, in 1999, the "bazaar" open-source and public software development method was published and advocated by Eric S. Raymond in Raymond, *The Cathedral and the Bazaar: Musings on Linux and Open Source by an Accidental Revolutionary*.

[70] "Cyber attacks lack the catastrophic dimensions of nuclear weapons attacks, and attribution is

signaling, and the ability to manage the escalation ladder.[71] In such a view, it may be counterproductive for nation-states to invest excessively in defensive cybersecurity. For the sake of this report, we leave such concerns aside, operating on the premise that improving defensive cybersecurity is overall positive for society.

---

more difficult, but inter-state deterrence still exists. Even when the source of an attack can be successfully disguised under a 'false flag,' other governments may find themselves sufficiently entangled in interdependent relationships that a major attack would be counterproductive." Nye, "Cyber Power."
[71] Valeriano, Jensen, and Maness, "Cyber Strategy: The Evolving Character of Power and Coercion."

# Acknowledgements

We would like to thank the UK AI Security Institute for supporting this work. In addition, we are grateful to the following people for providing valuable feedback and insights: Caleb Withers, Gil Gekker, John Halstead, Kyle Kilian, Matthew van der Merwe, Nikhil Mulani, Omer Nevo, Steven Adler, and Zoe Williams. All remaining errors are our own. Thanks also to Shane Coburn for copyediting support, Sherry Yang for graphics design, and Sarina Wong for assistance with the online publication of this report.